\documentclass[article]{jss}
\usepackage{amsmath,amssymb,amsfonts,amsthm,bbm,mathrsfs,natbib}
\usepackage{cancel,sectsty,color,enumitem,array,lscape,hyperref,graphicx,comment,subfig,lscape,scalefnt,framed}
\usepackage[title, titletoc]{appendix}
\usepackage[onehalfspacing]{setspace}
\usepackage[justification=centering]{caption}

\author{Sebastian Calonico\\Columbia University\And 
	    Matias D. Cattaneo\\Princeton University\And
        Max H. Farrell\\University of Chicago}
\title{\pkg{nprobust}: Nonparametric Kernel-Based Estimation and Robust Bias-Corrected Inference}

\Plainauthor{Sebastian Calonico, Matias D. Cattaneo, Max H. Farrell} 
\Plaintitle{nprobust: Nonparametric Kernel-Based Estimation and Robust Bias-Corrected Inference} 
\Shorttitle{\pkg{nprobust}: Nonparametric Kernel-Based Methods} 

\Abstract{
Nonparametric kernel density and local polynomial regression estimators are very popular in Statistics, Economics, and many other disciplines. They are routinely employed in applied work, either as part of the main empirical analysis or as a preliminary ingredient entering some other estimation or inference procedure. This article describes the main methodological and numerical features of the software package \pkg{nprobust}, which offers an array of estimation and inference procedures for nonparametric kernel-based density and local polynomial regression methods, implemented in both the  \proglang{R} and \proglang{Stata} statistical platforms. The package includes not only classical bandwidth selection, estimation, and inference methods \citep{Wand-Jones_1995_Book,Fan-Gijbels_1996_Book}, but also other recent developments in the statistics and econometrics literatures such as robust bias-corrected inference and coverage error optimal bandwidth selection \citep*{Calonico-Cattaneo-Farrell_2018_JASA,Calonico-Cattaneo-Farrell_2019_wp}. Furthermore, this article also proposes a simple way of estimating optimal bandwidths in practice that always delivers the optimal mean square error convergence rate regardless of the specific evaluation point, that is, no matter whether it is implemented at a boundary or interior point. Numerical performance is illustrated using an empirical application and simulated data, where a detailed numerical comparison with other \proglang{R} packages is given.

\begin{center}
	Last update: \today
\end{center}}

\Plainkeywords{kernel-based nonparametrics, bandwidth selection, bias correction, robust inference, R, Stata} 

\Address{
	Sebastian Calonico\\
    Mailman School of Public Health\\
    Columbia University\\
    New York, NY, USA\\
	E-mail: \email{scalonico@gmail.com}\\
    URL: \url{https://sites.google.com/site/scalonico/}\\

	Matias D. Cattaneo\\
	Department of Operations Research and Financial Engineering\\
	Princeton University\\
	Princeton, NJ, USA\\
	E-mail: \email{cattaneo@princeton.edu}\\
	URL: \url{https://cattaneo.princeton.edu/}\\
	
	Max H. Farrell\\
	Booth School of Business\\
	University of Chicago\\
	Chicago, IL, USA\\
	E-mail: \email{max.farrell@chicagobooth.edu}\\
	URL: \url{http://faculty.chicagobooth.edu/max.farrell/}
}

\usepackage{listings}
\definecolor{codegreen}{rgb}{0,0.6,0}
\definecolor{codegray}{rgb}{0.5,0.5,0.5}
\definecolor{codepurple}{rgb}{0.58,0,0.82}
\definecolor{backcolour}{rgb}{0.95,0.95,0.95}
 
\lstdefinestyle{mystyle}{
	language=R,
    backgroundcolor=\color{backcolour},   
    commentstyle=\color{codegreen},
    numberstyle=\tiny\color{codegray},
    stringstyle=\color{codepurple},
    basicstyle=\footnotesize\ttfamily,
    breakatwhitespace=false,         
    breaklines=true,                 
    captionpos=b,                    
    keepspaces=true,                 
    numbers=none,                    
    numbersep=5pt,                  
    showspaces=false,                
    showstringspaces=false,
    showtabs=false,                  
    tabsize=2
}
\theoremstyle{definition}

\allowdisplaybreaks[4]

	\DeclareMathOperator*{\argmin}{arg\,min}

    \DeclareMathOperator*{\diag}{diag}

	\renewcommand{\E}{\mathbb{E}}
	\newcommand{\V}{\mathbb{V}}
	\renewcommand{\P}{\mathbb{P}}

	\newcommand{\h}{h}
	
	\newcommand{\US}{\mathtt{us}}
	\newcommand{\BC}{\mathtt{bc}}
	\newcommand{\RBC}{\mathtt{rbc}}
	\newcommand{\MSE}{\mathtt{mse}}
	\newcommand{\IMSE}{\mathtt{imse}}
	\newcommand{\CE}{\mathtt{ce}}

	\newcommand{\bb}{\mathbf{b}}

	\newcommand{\bbeta}{\boldsymbol{\beta}}

	\newcommand{\be}{\mathbf{e}}

	\newcommand{\br}{\mathbf{r}}
	\newcommand{\bR}{\mathbf{R}}
	\newcommand{\bW}{\mathbf{W}}

	\newcommand{\bY}{\mathbf{Y}}

	\newcommand{\G}{\boldsymbol{\Gamma}}
	\newcommand{\Gp}{\G_p}
	\newcommand{\Gq}{\G_q}

	\renewcommand{\L}{\boldsymbol{\Lambda}}
	\newcommand{\Lp}{\L_p}

	\newcommand{\bSig}{\boldsymbol{\Sigma}}

	\newcommand{\bXi}{\boldsymbol{\Xi}}

	\setenumerate[1]{label*={\bf(\alph*)}}

\defcitealias{Calonico-Cattaneo-Farrell_2018_JASA}{CCF}

\begin{document}

\section[Introduction]{Introduction}\label{sec:intro}

Nonparametric kernel-based density and local polynomial regression methods are popular in Statistics, Economics, and many other disciplines, as they are routinely employed in empirical work. Correct implementation of these nonparametric procedures, however, often requires subtle (but important) methodological and numerical decisions. In this article, we give a comprehensive discussion of the software package \pkg{nprobust}, which is available for both \proglang{R} and \proglang{Stata}, and offers modern statistical methods for bandwidth selection, estimation, and inference in the context of kernel density and local polynomial regression fitting. Furthermore, this article also presents a novel bandwidth selection approach, which always delivers the optimal rate of convergence no matter the evaluation point under consideration (i.e., the bandwidth methodology is ``boundary rate adaptive''), and therefore is the default method used in the package. All the methods are fully data-driven and are implemented paying particular attention to finite sample properties as well as to recent theoretical developments in the statistics and econometrics literatures.

The package \pkg{nprobust} implements three types of statistical procedures for both density and local polynomial kernel smoothing. First, it provides several data-driven optimal bandwidth selection methods, specifically tailored to both (\textit{i}) point estimation, based on both mean square error (MSE) and integrated MSE (IMSE) approximations, and (\textit{ii}) confidence interval estimation or, equivalently, two-sided hypothesis testing, based on coverage error (CE) approximations via valid higher-order Edgeworth expansions. Second, the package implements point and variance estimators for density estimation at an interior point, and regression function estimation both at interior and boundary points. Finally, the package offers conventional and robust bias-corrected inference for all settings considered: density at interior point and regression function at both interior and boundary points. Some of these implementations build on classical results in the nonparametrics literature such as those reviewed in \citet{Wand-Jones_1995_Book} and \citet{Fan-Gijbels_1996_Book}, while others build on more recent developments on nonparametrics \citep*[][and references therein]{Calonico-Cattaneo-Farrell_2018_JASA,Calonico-Cattaneo-Farrell_2019_wp} and/or new practical ideas explained below. See also \citet{Ruppert-Wand-Carroll_2009_book} for semi/nonparametric applications in statistics, and \citet{Li-Racine_2007_Book} for semi/nonparametric applications in econometrics.

Section \ref{sec:overview} below gives an heuristic introduction to the main methodological ideas and results underlying the package implementation, avoiding technical details as much as possible, and focusing instead on the big picture overview of the methods available. For a technical discussion of the methods, see \citet*{Calonico-Cattaneo-Farrell_2018_JASA,Calonico-Cattaneo-Farrell_2019_wp} and their supplemental appendices. Related software implementing partitioning-based nonparametric methods, including binscatter, splines, and wavelets, are discussed in \citet{Cattaneo-Farrell-Feng_2019_lspartition} and \citet{Cattaneo-Crump-Farrell-Feng_2019_Stata}.

Unlike most other kernel smoothing implementations available in \proglang{R} and \proglang{Stata}, the package \pkg{nprobust} has two distinctive features, in addition to offering several new statistical procedures currently unavailable. First, all objects entering the different statistical procedures are computed using pre-asymptotic formulas, which has been shown to deliver superior estimation and inference. For example, when implemented in pre-asymptotic form, local polynomial regression inference at a boundary point exhibits higher-order boundary carpentry, while this important distributional feature is not present when asymptotic approximations are used instead for Studentization purposes. Second, also motivated by pre-asymptotic approximations and better finite sample properties, all inference procedures are constructed using both heteroskedasticity consistent (HC) variance estimators and cluster robust variance estimators. This approach departs from those employing homoskedasticity consistent variance estimators, which are implemented in most (if not all) packages currently available, because it has been found theoretically and in simulations to deliver more accurate inference procedures. The package \pkg{nprobust} implements HC variance estimators in two distinct ways: (i) using plug-in estimated residuals as usually done in least squares estimation (see \citet{Long-Ervin_2000_AS}, \citet{MacKinnon_2012_BookCh}, and references therein), and (ii) using nearest neighbor (NN) residuals (motivated by \citet{Muller-Stadtmuller_1987_AoS} and \citet{Abadie-Imbens_2008_AdES}). In particular, both the HC$3$ and NN variance estimators have been found to perform quite well in simulations. Cluster robust variance estimation is implemented using either plug-in estimated residuals with degrees-of-freedom adjustment or NN residuals.

In this article, we focus the discussion on the \proglang{R} version of the package \pkg{nprobust}, which makes use of the packages \pkg{ggplot2} \citep{ggplot2} and \pkg{Rcpp} \citep{Rcpp}. Nevertheless, we remark that the \textit{exact} same functionalities are available in its \proglang{Stata} version. The package \pkg{nprobust} is organized through five general purpose functions (or commands in \proglang{Stata}):

\begin{itemize}
	
	\item \code{lprobust()}: this function implements estimation and inference procedures for kernel-based local polynomial regression at both interior and boundary points in an unified way. This function takes the bandwidth choices as given, and implements point and variance estimation, bias correction, conventional and robust bias-corrected confidence intervals, and related inference procedures. All polynomial orders are allowed and the implementation automatically adapts to boundary points, given the choice of bandwidth(s). When the bandwidth(s) are not specified, the companion function \code{lpbwselect()} is used to select the necessary bandwidth(s) in a fully data-driven, automatic fashion, while taking into account whether the evaluation point is an interior or boundary point.
	
	\item  \code{lpbwselect()}: this function implements bandwidth selection for kernel-based local polynomial regression methods at both interior and boundary points. Six distinct bandwidth selectors are available: direct plug-in (DPI) and rule-of-thumb (ROT) implementations of MSE-optimal and IMSE-optimal choices (four alternatives), and DPI and ROT implementations of the CE-optimal choice (two alternatives).
	
	\item \code{kdrobust()}: this function implements estimation and inference procedures for kernel density estimation at an interior point. Like the function \code{lprobust()}, this function also takes the bandwidth choices as given and implements point and variance estimation, bias correction, conventional and robust bias-corrected confidence intervals, and related inference procedures. When the bandwidth(s) are not specified, the companion function \code{kdbwselect()} is used. This function cannot be used for density estimation at boundary points; see \citet*{Cattaneo-Jansson-Ma_2019_lpdensity} for boundary-adaptive kernel density estimation methods.
	
	\item \code{kdbwselect()}: this function implements bandwidth selection for kernel density estimation at an interior point. Mimicking the structure and options of the function \code{lpbwselect()}, the same type of bandwidth selectors are available.
	
	\item \code{nprobust.plot}: this function is used as wrapper for plotting results, and is only available in \proglang{R} because it builds on the package \code{ggplot2}. Analogous plotting capabilities are available in \proglang{Stata}, as we illustrate in our companion replication files for the latter platform.
	
\end{itemize}

We also present detailed numerical evidence on the performance of the package \pkg{nprobust}. First, in Section \ref{sec:ampapp}, we illustrate several of its main functionalities using the canonical dataset of \citet{Efron-Feldman_1991_JASA}. Second, in Section \ref{sec:simuls}, we investigate the finite sample performance of the package using a simulation study. We also compare its performance to four other popular \proglang{R} packages implementing similar nonparametric methods: see Table \ref{table:Rpkgs} for details. We find that the package \pkg{nprobust} outperforms the alternatives in most cases, and never underperforms in terms of bandwidth selection, point estimation, and inference. Finally, for comparability, the Appendix discusses the \proglang{R} syntax for all the packages considered.

Installation details, scripts in \proglang{R} and \proglang{Stata} replicating all numerical results, links to software repositories, and other companion information concerning the package \pkg{nprobust} can be found in the website: \url{https://sites.google.com/site/nppackages/nprobust/}.

\section[Overview of Methods and Implementation]{Overview of Methods and Implementation}\label{sec:overview}

This section offers a brief, self-contained review of the main methods implemented in the package \pkg{nprobust} for the case of local polynomial regression. Whenever possible, we employ exactly the same notation as in \citet*[][CCF hereafter]{Calonico-Cattaneo-Farrell_2018_JASA} to facilitate comparison and cross-reference with their lengthy supplemental appendix. See also \citet*{Calonico-Cattaneo-Farrell_2019_wp} for extensions to derivative estimation and other theoretical and practical results. The case of kernel density estimation at interior points is not discussed here to conserve space, but is nonetheless implemented in the package \pkg{nprobust}, as illustrated below. Further details can be found in the classical textbooks such as \citet{Wand-Jones_1995_Book} and \citet{Fan-Gijbels_1996_Book}, as well as in \citetalias{Calonico-Cattaneo-Farrell_2018_JASA} and references therein.

We assume that $(Y_i,X_i)$, with $i=1,2,\ldots,n$, is a random sample with $m(x)=\E[Y_i|X_i=x]$, or its derivative, begin the object of interest. The evaluation point $x$ can be an ``interior'' or a ``boundary'' point, and we will discuss this issue when relevant. Regularity conditions such as smoothness or existence of moments are omitted herein, but can be found in the references already given. We discuss how the point estimator, its associated bias-corrected point estimator, and their corresponding variance estimators are all constructed. We employ two bandwidths $h$ and $b$, where $h$ is used to construct the original point estimator and $b$ is used to construct the bias correction (robust bias-corrected inference allows for $h=b$). Finally, in our notation any object indexed by $p$ is computed with bandwidth $h$ and any object indexed by $q$ is computed with bandwidth $b$ and $q=p+1$. We set $\rho=h/b$ for future reference. When appealing to asymptotic approximations, limits are taken $h\to0$ and $b\to0$ as $n\to\infty$ unless explicitly stated otherwise. Finally, for any smooth function $g(x)$, we use the notation $g^{(\nu)}(x)=d^\nu g(x)/dx^\nu$ to denote its $\nu$-th derivative, with $g(x)=g^{(0)}(x)$ to save notation.

\subsection{Point Estimation}

The package \pkg{nprobust} implements fully data-driven, automatic kernel-based estimation and inference methods for the regression function $m(x)$ and its derivatives. Since kernel-based local polynomial regression estimators are consistent at all evaluation points, we consider both interior and boundary points simultaneously.

The classical local polynomial estimator of $m^{(\nu)}(x)$, $0\leq\nu\leq p$, is
\[\hat{m}^{(\nu)}(x) = \nu! \be_\nu' \hat{\bbeta}_p, \qquad\quad
\hat{\bbeta}_p = \argmin_{\bb \in \mathbb{R}^{p+1}} \sum_{i=1}^n ( Y_i - \br_p(X_i - x)'\bb)^2  K \left( \frac{X_i - x}{\h}\right),
\]
where, for a non-negative integer $p$, $\be_\nu$ is the $(\nu+1)$-th unit vector of $(p+1)$-dimension, $\br_p(u) = (1, u, u^2, \ldots, u^p)'$, and $K(\cdot)$ denotes a symmetric kernel with bounded support. See \citet{Fan-Gijbels_1996_Book} for further discussion and basic methodological issues. While \citetalias{Calonico-Cattaneo-Farrell_2018_JASA} restrict attention to $p$ odd, as is standard in the local polynomial literature, the package \pkg{nprobust} allows for any $p\geq0$. In particular, $p=0$ corresponds to the classical Nadaraya-Watson kernel regression estimator, while $p=1$ is the local linear estimator with the celebrated boundary carpentry feature. Some of the higher-order results reported in \citetalias{Calonico-Cattaneo-Farrell_2018_JASA} for local polynomial estimators may change when $p$ is even, as we discuss further below, but all the first-order properties remain valid and hence the package \pkg{nprobust} allows for any $p\geq0$. The package uses $p=1$ as default, in part because all the first- and higher-order results reported in \citet{Fan-Gijbels_1996_Book}, \citetalias{Calonico-Cattaneo-Farrell_2018_JASA}, and references therein apply. The package offers three different kernel choices: Epanechnikov, Triangular, and Uniform.

For any fixed sample size, the local polynomial estimator $\hat{m}^{(\nu)}(x)$ is a weighted least squares estimator, and hence can be written in matrix form as follows:  $\hat{m}^{(\nu)}(x) = \nu!\be_\nu'\Gp^{-1} \bR_p' \bW_p \bY / n$, where $\bY = (Y_1, \cdots, Y_n)'$, $\bR_p = [ \br_p( (X_1 - x) / \h), \cdots, \br_p( (X_n - x) / \h)]'$, $\bW_p = \diag(K((X_i - x)/\h)/\h: i = 1, \ldots, n)$, and $\Gp = \bR_p' \bW_p \bR_p/n$; here $\diag(a_i:i = 1, \ldots, n)$ denotes the $n\times n$ diagonal matrix constructed using $a_1, a_2, \cdots, a_n$. This representation is convenient to develop pre-asymptotic approximations for estimation and inference, as we discuss below.

\subsection{Bias and Variance}

The basic statistical properties of the nonparametric estimators $\hat{m}^{(\nu)}(x)$ are captured by their bias and variance. In pre-asymptotic form, that is, for fixed sample size and bandwidth choice, the (conditional) bias and variance of $\hat{m}^{(\nu)}(x)$ can be represented by
\[\mathscr{B}[\hat{m}^{(\nu)}(x)] \approx h^{p+1-\nu} \left(\mathcal{B}_1[\hat{m}^{(\nu)}(x)] + h\mathcal{B}_2[\hat{m}^{(\nu)}(x)]\right)
\quad\text{and}\quad
\mathscr{V}[\hat{m}^{(\nu)}(x)] = \frac{1}{nh^{1+2\nu}} \mathcal{V}[\hat{m}^{(\nu)}(x)],\]
respectively, where the objects $\mathcal{B}_1[\hat{m}^{(\nu)}(x)]$, $\mathcal{B}_2[\hat{m}^{(\nu)}(x)]$ and $\mathcal{V}[\hat{m}^{(\nu)}(x)]$ depend on observable quantities such as $n$, $h$ and $K(\cdot)$, and each depend on only one unknown quantity: $m^{(p+1)}(x)$, $m^{(p+2)}(x)$, and the heteroskedasticity function $\sigma^2(X_i)=\V[Y_i|X_i]$, respectively. Throughout the paper, $\approx$ means that the approximation holds for large samples in probability.

To be more specific, the pre-asymptotic variance $\mathcal{V}[\hat{m}^{(\nu)}(x)]$ takes the form:
\[\mathcal{V}[\hat{m}^{(\nu)}(x)]
= \frac{\h}{n} \nu!^2\be_\nu' \Gp^{-1} \bR_p' \bW_p {\bSig} \bW_p \bR_p \Gp^{-1} \be_\nu,\]
which is valid for any $\nu$ and $p$ and for interior and boundary points $x$, as this formula follows straightforwardly from linear least squares algebra, where $\bSig = \diag(\sigma^2(X_i): i = 1, \ldots, n)$. In practice, $\bSig$ is replaced by some ``estimator'', leading to heteroskedasticity consistent (HC) or cluster robust variance estimation for (weighted) linear least squares fitting.

The pre-asymptotic bias, on the other hand, is more complicated as it depends on whether $x$ is an interior or a boundary point, and whether $p-\nu$ is odd or even. To be concrete, define
\[\mathcal{B}_1[\hat{m}^{(\nu)}(x)] = \frac{\nu!}{(p+1)!} \be_\nu' \Gp^{-1} \L_{p} m^{(p+1)}(x),\qquad
\mathcal{B}_2[\hat{m}^{(\nu)}(x)] = \frac{\nu!}{(p+2)!} \be_\nu' \Gp^{-1} \L_{p+1} m^{(p+2)}(x),\]
where $\Lp = \bR_p' \bW_p [ ((X_1 - x)/\h)^{p+1}, \cdots, ((X_n - x)/\h)^{p+1}]'/n$. Then, if $p-\nu$ is odd \textit{or} $x$ is a boundary point, $\mathcal{B}_1[\hat{m}^{(\nu)}(x)]$ is the leading term of the bias (i.e., regardless of the evaluation point when $p-\nu$ is odd), and $h\mathcal{B}_2[\hat{m}^{(\nu)}(x)]$ is negligible in large samples. Alternatively, if $p-\nu$ is even \textit{and} $x$ is an interior point, then
\[\mathcal{B}_1[\hat{m}^{(\nu)}(x)] \approx h \mathcal{B}_{11}[\hat{m}^{(\nu)}(x)],\]
where the quantity $\mathcal{B}_{11}[\hat{m}^{(\nu)}(x)]$ is non-zero, and relatively easy to estimate using a pilot bandwidth and pre-asymptotic approximations. (It is, however, notationally cumbersome to state explicitly.) Therefore, when $p-\nu$ is even \textit{and} $x$ is an interior point, both $\mathcal{B}_1[\hat{m}^{(\nu)}(x)]$ and $h\mathcal{B}_2[\hat{m}^{(\nu)}(x)]$ are part of the leading bias. This distinction is important when it comes to (automatic) implementation, as the form of the bias changes depending on the evaluation point and regression estimator considered.

At a conceptual level, putting aside technical specifics on approximation and implementation, the bias and variance quantities given above play a crucial role in bandwidth selection, estimation, and inference. The package \pkg{nprobust} pays special attention to constructing robust estimates of these quantities, with good finite sample and boundary properties.

\subsection{Mean Square Error and Bandwidth Selection}

In order to implement the point estimator $\hat{m}^{(\nu)}(x)$, a choice of bandwidth $h$ is needed. Following the classical nonparametric literature, the package \pkg{nprobust} employs pointwise and integrated MSE approximations for the point estimator to compute the corresponding MSE-optimal or IMSE-optimal bandwidth for point estimation.

The pointwise MSE approximation is
\[\mathsf{MSE}[\hat{m}^{(\nu)}(x)]\approx \mathscr{B}[\hat{m}^{(\nu)}(x)]^2+\mathscr{V}[\hat{m}^{(\nu)}(x)],\]
while the integrated MSE approximation is
\[\int\mathsf{MSE}[\hat{m}^{(\nu)}(x)]w(x)dx\approx \int\mathscr{B}[\hat{m}^{(\nu)}(x)]^2w(x)dx+\int\mathscr{V}[\hat{m}^{(\nu)}(x)]w(x)dx\]
with $w(x)$ a weighting function.

\subsubsection{Case 1: $p-\nu$ odd}
Employing the approximations above, the MSE-optimal bandwidth is
\[h_\MSE = \argmin_{h>0}\left|h^{2(p+1-\nu)} \left(\mathcal{B}_1[\hat{m}^{(\nu)}(x)]\right)^2
+ \frac{1}{nh^{1+2\nu}} \mathcal{V}[\hat{m}^{(\nu)}(x)] \right|,\]
which, in this case, has closed form solution given by
\[h_\MSE = \left(\frac{(1+2\nu)\mathcal{V}[\hat{m}^{(\nu)}(x)]}
{2(p+1-\nu)(\mathcal{B}_1[\hat{m}^{(\nu)}(x)])^2}\right)^{1/(2p+3)} n^{-1/(2p+3)}.
\]
Here we abuse notation slightly: the quantities $\mathcal{V}[\hat{m}^{(\nu)}(x)]$ and $\mathcal{B}[\hat{m}^{(\nu)}(x)]$ are taken as fixed and independent of $n$ and $h$, that is, denoting their probability limits as opposed to pre-asymptotic, as originally defined. Nevertheless, we abuse notation because in the end they will be approximated using their pre-asymptotic counterparts with a preliminary/pilot bandwidth.

Similarly, the integrated MSE approximation when $p-\nu$ is odd leads to the IMSE-optimal bandwidth
\[h_\IMSE = \argmin_{h>0}\left|h^{2(p+1-\nu)} \int\left(\mathcal{B}_1[\hat{m}^{(\nu)}(x)]\right)^2w(x)dx
+ \frac{1}{nh^{1+2\nu}} \int\mathcal{V}[\hat{m}^{(\nu)}(x)]w(x)dx \right|,\]
which also has a closed form solution given by
\[h_\IMSE = \left(\frac{(1+2\nu)\int\mathcal{V}[\hat{m}^{(\nu)}(x)]w(x)dx}
{2(p+1-\nu)\int(\mathcal{B}_1[\hat{m}^{(\nu)}(x)])^2w(x)dx}\right)^{1/(2p+3)} n^{-1/(2p+3)}.
\]

These MSE-optimal and IMSE-optimal bandwidth selectors are valid for $p-\nu$ odd, regardless of whether $x$ is a boundary or an interior point. Implementation of these selectors is not difficult, as the bias and variance quantities are taken to be pre-asymptotic and hence most parts are already in estimable form. For example, we have the following easy to implement bias estimator:
\[\hat{\mathcal{B}}[\hat{m}^{(\nu)}(x)] = \frac{\nu!}{(p+1)!} \be_\nu' \Gp^{-1} \L_p \hat{m}^{(p+1)}(x),\]
where $\Gp$ and $\Lp$ are pre-asymptotic and hence directly estimable once a preliminary/pilot bandwidth is set, and $\hat{m}^{(p+1)}(x)$ is simply another local polynomial fit of order $q=p+1$. Similarly, the fixed-$n$ variance estimator is given by:
\[\hat{\mathscr{V}}[\hat{m}^{(\nu)}(x)]=\frac{1}{nh^{1+2\nu}}\nu!^2\be_\nu' \Gp^{-1} \bR_p' \bW_p \hat{\bSig} \bW_p \bR_p \Gp^{-1} \be_\nu,\]
where all the objects are directly computable from the data and valid for all $p-\nu$ (odd or even) and all evaluation points (interior or boundary), provided that $\hat{\bSig}$ is chosen appropriately. Following \citetalias{Calonico-Cattaneo-Farrell_2018_JASA}, the package \pkg{nprobust} allows for five different $\hat{\bSig}$ under unknown conditional heteroskedasticity: HC0, HC1, HC2, HC3, and NN. The first four choices employ weighted estimated residuals, motivated by a weighted least squares interpretation of the local polynomial estimator, while the last choice uses a nearest neighbor approach to constructing residuals. Details underlying these choices, and their properties in finite and large samples, are discussed in \citetalias{Calonico-Cattaneo-Farrell_2018_JASA} and its supplemental appendix. In addition, the package \pkg{nprobust} allows for two cluster robust choices for $\hat{\bSig}$: plug-in residuals with degrees-of-freedom adjustment (similar to HC1) and NN-cluster residuals (similar to NN).

As mentioned in the introduction, the outlined pre-asymptotic approach for bias and variance estimation is quite different from employing the usual asymptotic approximations, which is the way most implementations of kernel smoothing are done in both \proglang{R} and \proglang{Stata}. Importantly, using valid higher-order Edgeworth expansions, \citetalias{Calonico-Cattaneo-Farrell_2018_JASA} established formally that using pre-asymptotic approximations for bias and variance quantities delivers demonstrably superior distributional approximations, and hence better finite sample performance.

\subsubsection{Case 2: $p-\nu$ even}
This case requires additional care. If $x$ is a boundary point, then the above formulas are valid and the same MSE-optimal and IMSE-optimal bandwidth selectors can be used. However, when $x$ is an interior point \textit{and} $p-\nu$ even, the leading bias formulas change and so do the corresponding bandwidth selectors. Therefore, to account for this issue in an automatic way, since in any given application it is not possible to cleanly distinguish between interior and boundary evaluation points, the default bandwidth selector implemented in the package \pkg{nprobust} optimizes the pre-asymptotic full bias approximation, leading to the following MSE-optimal and IMSE-optimal bandwidth selectors
\[h_\MSE = \argmin_{h>0}\left|h^{2(p+1-\nu)} \left(\mathcal{B}_1[\hat{m}^{(\nu)}(x)] + h\mathcal{B}_2[\hat{m}^{(\nu)}(x)]\right)^2
+ \frac{1}{nh^{1+2\nu}} \mathcal{V}[\hat{m}^{(\nu)}(x)] \right|\]
and
\[h_\IMSE = \argmin_{h>0}\left|h^{2(p+1-\nu)} \int\left(\mathcal{B}_1[\hat{m}^{(\nu)}(x)] 
+ h\mathcal{B}_2[\hat{m}^{(\nu)}(x)]\right)^2w(x)dx
+ \frac{1}{nh^{1+2\nu}} \int\mathcal{V}[\hat{m}^{(\nu)}(x)]w(x)dx \right|,\]
respectively.

These bandwidth selectors for local polynomial regression with $p-\nu$ even automatically adapt to the evaluation point in terms of the resulting convergence rate, i.e. the bandwidth choice will deliver point estimates with the MSE- or IMSE-optimal convergence rate, but cannot be given in closed form. Furthermore, the package also offers the possibility of explicitly treating all points $x$ as interior points when $p-\nu$ even, in which case the estimator of the term $\mathcal{B}_1[\hat{m}^{(\nu)}(x)]$ is replaced by an estimator of the term $h\mathcal{B}_{11}[\hat{m}^{(\nu)}(x)]$, leading to the standard bandwidth selectors, which are more cumbersome to state but can be computed in closed form.

\subsubsection{Boundary Adaptive Bandwidth Selection}

The function \code{lpbwselect()} implements a DPI estimate of the MSE-optimal and IMSE-optimal bandwidth selectors discussed previously, taking into account whether $p-\nu$ is even or odd and whether $x$ is interior or boundary point. These bandwidth choices depend on $p$, $\nu$, $K(\cdot)$, and the preliminary bias and variance estimators. For the bias quantity, the preliminary estimator is computed using a nonparametric point estimators for the unknown quantities $m^{(p+1)}$, $m^{(p+2)}$, etc., depending on the case of local polynomial regression considered, but in all cases preliminary quantities are estimated using MSE-optimal bandwidth choices. For the variance quantity, a choice of ``estimated'' residual is used, as already discussed, and a preliminary rule-of-thumb bandwidth is employed to construct the remaining pre-asymptotic objects. The integrated versions are computed by taking a sample average of the pointwise bias and variance quantities over the evaluation points selected by the user. Finally, the function \code{lpbwselect()} also implements ROT versions of the MSE-optimal and IMSE-optimal bandwidth selectors for completeness.

All the proposed and implemented bandwidth selection methods outlined above are rate adaptive in the sense that the selected bandwidth always exhibits the (I)MSE-optimal convergence rate, regardless of the polynomial order ($p$), the derivative order ($\nu$), and the evaluation point ($x$) considered. Furthermore, in the case of $p-\nu$ odd or $x$ a boundary point, the plug-in estimated constant entering the bandwidth selectors will also be consistent for the corresponding (I)MSE-optimal population constant. On the other hand, in the case of $p-\nu$ even and $x$ an interior point, the resulting bandwidth selectors will not estimate consistently the complete bias constant (only one of the two constants is estimated consistently). To resolve the later issue in cases where the user is considering only interior points, the function \code{lpbwselect()} has the option \code{interior} which implements bandwidth selection for interior points only in all cases.

\subsection{Optimal Point Estimation}

When implemented using the (pointwise) MSE-optimal bandwidth, the regression function estimator $\hat{m}^{(\nu)}(x)$ is an MSE-optimal point estimator, in the sense that it minimizes the pointwise asymptotic mean square error objective function. Sometimes, researchers (and other software packages) implementing kernel smoothing methods prefer to optimize the bandwidth choice in a global sense and thus focus on optimal bandwidths targeting the IMSE, as presented above, or some form of cross-validation objective function. These alternative bandwidth selectors also lead to optimal point estimators in some sense. Regardless of the specific objective function considered, and hence bandwidth rule used, the resulting bandwidth selectors exhibit the same rate of decay as $n \to \infty$ (albeit different constants) in all cases: for example, compare $h_\MSE$ and $h_\IMSE$ given above in close form for the case of $p-\nu$ odd.

The package \pkg{nprobust} implements only plug-in bandwidth selectors for both local (pointwise) and global (integrated) bandwidth choices, because such rules tends to have better finite sample properties. Cross-validation methods are not implemented due to potential numerical instabilities, and potentially low convergence rates of the resulting bandwidth choices. Nevertheless, the package can of course be used to construct cross-validation bandwidth choices manually.

\subsection{Conventional and Robust Bias-Corrected Inference}

Given a choice of bandwidth, \textit{conventional} inference in nonparametric kernel-based estimation employs a Gaussian distributional approximation for the usual Wald-type statistic. To be more precise, standard asymptotic results give
\[ T_\US(x) = \frac{\hat{m}^{(\nu)}(x) - m^{(\nu)}(x)}{\sqrt{\hat{\mathscr{V}}[\hat{m}^{(\nu)}(x)]}}
\rightsquigarrow \mathcal{N}(0,1),
\]
where $\hat{\mathscr{V}}[\hat{m}^{(\nu)}(x)]$ denotes an estimator of $\mathscr{V}[\hat{m}^{(\nu)}(x)]$, already discussed above, and $\rightsquigarrow$ denotes convergence in distribution. Using this classical result, we obtain the nominal $(1-\alpha)$-percent \textit{conventional} symmetric confidence intervals:
\[ I_\US(x) = \left[ \; \hat{m}^{(\nu)}(x) - \Phi_{1-\alpha/2} \sqrt{\hat{\mathscr{V}}[\hat{m}^{(\nu)}(x)]} \; , \;
\hat{m}^{(\nu)}(x) - \Phi_{\alpha/2} \sqrt{\hat{\mathscr{V}}[\hat{m}^{(\nu)}(x)]} \;\right],
\]
where $\Phi_{u}=\Phi^{-1}(u)$ with $\Phi(u)$ denoting the standard normal cumulative distribution function.

However, this standard distributional approximation result crucially requires undersmoothing ($\US)$ of the bandwidth employed: the confidence interval $I_\US(x)$ will not have correct coverage when any of the mean square error optimal bandwidths discussed previously are used. The main reason is that, when $h_\MSE$ or $h_\IMSE$ (or a cross-validation bandwidth) is employed to construct the local polynomial point estimator $\hat{m}^{(\nu)}(x)$, then $T_\US(x)\rightsquigarrow\mathcal{N}(\mathfrak{b},1)$, where $\mathfrak{b}$ denotes a non-vanishing asymptotic bias. Furthermore, if a larger bandwidth than an MSE optimal is used, then the distributional approximation fails altogether.

This observation is very important because most (if not all) available kernel smoothing implementations in \proglang{R} and \proglang{Stata} employ a mean square error optimal bandwidth for both point estimation and inference, which means that the corresponding inference results are incorrect. One solution to this problem is to undersmooth the bandwidth, that is, to employ a smaller bandwidth for conducting inference and constructing confidence intervals. However, while theoretically valid in large samples, this approach does not typically perform well in applications, leads to power losses, and requires employing different observations for estimation and inference purposes, all important drawbacks for empirical work.

Motivated in part by the above issues, \citetalias{Calonico-Cattaneo-Farrell_2018_JASA} developed new inference methods based on nonparametric bias correction, which they termed Robust Bias Correction (RBC). This approach is based on traditional plug-in bias correction, as an alternative to undersmoothing, but employs a different variance estimator for Studentization purposes (when compared to those discussed in textbooks). The key underlying idea is that because the bias is estimated when conducting bias correction, the variance estimator should be change to account for the additional variability introduced. This leads to a new test statistic, where both the centering (bias) and the scale (variance) have been adjusted, but the same bandwidth can be used. To be more specific, we have
\[ T_\RBC(x) = \frac{\hat{m}_\BC^{(\nu)}(x) - m^{(\nu)}(x)}{\sqrt{\hat{\mathscr{V}}[\hat{m}_\BC^{(\nu)}(x)]}}
\rightsquigarrow \mathcal{N}(0,1), \qquad
\hat{m}_\BC^{(\nu)}(x) = \hat{m}^{(\nu)}(x) - \hat{\mathscr{B}}[\hat{m}^{(\nu)}(x)],
\]
where $\hat{\mathscr{V}}[\hat{m}_\BC^{(\nu)}(x)]$ denotes a variance estimator of $\mathscr{V}[\hat{m}_\BC^{(\nu)}(x)]$ and $\hat{\mathscr{B}}[\hat{m}^{(\nu)}(x)]$ denotes a bias estimator (i.e., an estimate of $\mathscr{B}[\hat{m}^{(\nu)}(x)]$). The bias-corrected point estimator $\hat{m}_\BC^{(\nu)}(x)$ and its fixed-$n$ variance estimator $\hat{\mathscr{V}}[\hat{m}_\BC^{(\nu)}(x)]$ are, respectively,
\[\hat{m}_\BC^{(\nu)}(x) = \nu!\be_\nu'\Gp^{-1} \bXi_{p,q} \bY / n \quad\text{and}\quad
\hat{\mathscr{V}}[\hat{m}_\BC^{(\nu)}(x)]=\frac{1}{nh^{1+2\nu}}\nu!^2\be_\nu' \Gp^{-1} \bXi_{p,q} \hat{\bSig} \bXi_{p,q}' \Gp^{-1} \be_\nu,
\]
where $\bXi_{p,q} = \bR_p' \bW_p  -  \rho^{p+1} \Lp \be_{p+1}' \Gq^{-1} \bR_q' \bW_q$. As explained in \citetalias{Calonico-Cattaneo-Farrell_2018_JASA}, this formula emerges from employing $q$-th order local polynomial with bandwidth $b$ to estimate the leading higher-order derivative featuring in the fixed-$n$ bias approximation for the estimator $\hat{m}^{(\nu)}(x)$. Finally, in this case, $\hat{\bSig}$ is also computed using any of the five HC variance estimators (HC0, HC1, HC2, HC3, NN) or the two cluster-robust versions mentioned above. The estimator $\hat{\mathscr{B}}[\hat{m}^{(\nu)}(x)]$ was already introduced above for bandwidth selection, but the estimator $\hat{\mathscr{V}}[\hat{m}_\BC^{(\nu)}(x)]$ is new because it accounts for the variance of $\hat{m}^{(\nu)}(x)$, the variance of $\hat{\mathscr{B}}[\hat{m}^{(\nu)}(x)]$, and the covariance between these terms.

The RBC test statistic can be constructed with any of the (I)MSE optimal bandwidths, including the ones introduced above, and the standard Gaussian approximation remains valid. This result leads to the nominal $(1-\alpha)$-percent \textit{robust bias-corrected} symmetric confidence intervals:
\[ I_\RBC(x) = \left[\; \hat{m}_\BC^{(\nu)}(x) - \Phi_{1-\alpha/2} \sqrt{\hat{\mathscr{V}}[\hat{m}_\BC^{(\nu)}(x)]} \; , \;
\hat{m}_\BC^{(\nu)}(x) - \Phi_{\alpha/2} \sqrt{\hat{\mathscr{V}}[\hat{m}_\BC^{(\nu)}(x)]} \;\right].
\]

The package \pkg{nprobust} implements both conventional and RBC inference for local polynomial estimators, at interior and boundary points. \citetalias{Calonico-Cattaneo-Farrell_2018_JASA} also show that the same ideas and results apply to density estimation at an interior point, and the package also implements those results.

The function \code{lprobust()} implements the point estimators, variance estimators, conventional inference, and robust bias-corrected inference discussed above, taking into account whether the evaluation point $x$ is near the boundary or not, and employing all pre-asymptotic approximations for estimation of bias and variance. Two bandwidths are used: $h$ for constructing the main point estimator and related quantities, and $b$ for constructing the higher-order derivative entering the bias correction term. As discussed in \citetalias{Calonico-Cattaneo-Farrell_2018_JASA}, setting $h=b$ is allowed by the theory, and leads to a simple method for bias correction. This is the default in \pkg{nprobust}.

\subsection{Coverage Error and Bandwidth Selection}

An (I)MSE-optimal bandwidth can be used to construct optimal point estimators but cannot be used directly to conduct inference or form confidence intervals, in the conventional way, since a first-order bias renders the standard Gaussian approximation invalid as discussed above. Robust bias-corrected inference allows the use of an (I)MSE-optimal bandwidth to form a test statistic with a standard Gaussian distribution in large samples, thereby leading to valid confidence intervals and hypothesis tests.

However, while \emph{valid}, employing an (I)MSE-optimal bandwidth to form the robust bias-corrected statistic may not be \emph{optimal} from a distributional or inference perspective. \citetalias{Calonico-Cattaneo-Farrell_2018_JASA} (for $\nu=0$) and \citet*{Calonico-Cattaneo-Farrell_2019_wp} (for $\nu \geq 0$) study this problem formally using valid Edgeworth expansions and show that indeed the (I)MSE-optimal bandwidth is suboptimal in general, even after robust bias-correction, because it is too ``large'' (i.e., it decays to zero too slowly) relative to the optimal choice. A very important exception is $p=1$, $\nu=0$, with $x$ an interior point, in which case the MSE-optimal bandwidth does have the CE-optimal rate of convergence. 

Consider first the case of local polynomial regression with $p-\nu$ odd and robust bias correction inference. The asymptotic coverage of the robust bias-corrected confidence interval estimator $I_\RBC(x)$ is:
\[ \P[m^{(\nu)}(x) \in I_\RBC(x)] \approx 1-\alpha + \mathsf{CE}[\hat{m}_\BC^{(\nu)}(x)],\]
where the higher-order coverage error is
\[ \mathsf{CE}[\hat{m}_\BC^{(\nu)}(x)] = \frac{1}{nh} \mathcal{E}_{1,\nu}(x)
+ nh^{2p+5} (\mathcal{E}_{2,\nu}(x) + h\mathcal{E}_{3,\nu}(x))^2
+ h^{p+2} (\mathcal{E}_{4,\nu}(x) + h\mathcal{E}_{5,\nu}(x)).
\]
This representation follows from Corollary 5 in \citetalias{Calonico-Cattaneo-Farrell_2018_JASA} and results in \citet*{Calonico-Cattaneo-Farrell_2019_wp} (see also their supplemental appendices), and is written in a way that is adaptive to the evaluation point $x$, that is, it is asymptotically valid for both interior and boundary points. Thus, a CE-optimal bandwidth for implementing confidence intervals $I_\RBC(x)$, with $p$ odd, is:
\[h_\CE = \argmin_{h>0}\left|\frac{1}{h} \mathcal{E}_{1,\nu}(x)
+ nh^{2p+5} (\mathcal{E}_{2,\nu}(x) + h\mathcal{E}_{3,\nu}(x))^2
+ h^{p+2} (\mathcal{E}_{4,\nu}(x) + h\mathcal{E}_{5,\nu}(x))\right|.
\]
Like in the case of (I)MSE-optimal bandwidth selection, a plug-in CE-optimal bandwidth estimator is obtained by replacing $\mathcal{E}_{k,\nu}(x)$ with consistent estimators $\hat{\mathcal{E}}_{k,\nu}(x)$, $k=1,2,3,4,5$. The bandwidth selector $h_\CE$, and its data-driven implementation, can be used to construct robust bias-corrected confidence intervals $I_\RBC(x)$ with CE-optimal properties, that is, with minimal coverage error in large samples. Because the constants entering the coverage error are estimated consistently, the resulting CE-optimal bandwidth choice corresponds to a DPI bandwidth selector, which is available only for $p$ odd (and, of course, robust bias-corrected inference). For $p$ even, only the ROT CE-optimal bandwidth selector discussed below is currently available.

As discussed in the supplemental appendix of \citetalias{Calonico-Cattaneo-Farrell_2018_JASA}, it is always possible to construct a simple and intuitive ROT implementation of the CE-optimal bandwidth selector by rescaling any MSE-optimal bandwidth choice:
\[h_\CE = \begin{cases}
n^{-\frac{p}{(2p+3)(p+3)}} \; h_\MSE    & \quad\text{for $p$ odd} \\
n^{-\frac{p+2}{(2p+5)(p+3)}} \; h_\MSE  & \quad\text{for $p$ even}
\end{cases}\]
with $h_\MSE$ denoting the previously discussed MSE-optimal bandwidth for the point estimator $\hat{m}(x)$. This alternative bandwidth choice is available for all $p$, and is implemented using the DPI MSE-optimal bandwidth choice for $h_\MSE$ with $p$ denoting the polynomial order used construct the local polynomial regression point estimate. See the supplemental appendix of \citetalias{Calonico-Cattaneo-Farrell_2018_JASA} for more details, and an alternative ROT bandwidth choice that could be used in some specific cases.

\section[Empirical Illustration]{Empirical Illustration}\label{sec:ampapp}

This section showcases some of the features of the package \pkg{nprobust} employing the canonical dataset of \citet{Efron-Feldman_1991_JASA}, which corresponds to a randomized clinical trial evaluating the efficacy of cholestyramine for reducing cholesterol levels. The dataset used herein consists of five variables (\texttt{t}, \texttt{chol1}, \texttt{chol2}, \texttt{cholf}, \texttt{comp}) for $n = 337$ observations in a placebo-controlled double-blind clinical trial, with $172$ units assigned to treatment and $165$ units assigned to control. As mentioned previously, we focus exclusively on the \proglang{R} implementation of the package only for concreteness, but reiterate that the companion \proglang{Stata} implementation offers exact same features. In fact, we do provide a \proglang{Stata} do-file replicating all the results discussed in this section.

From within \proglang{R}, the latest version of \pkg{nprobust} can be installed from CRAN (The Comprehensive R Archive Network) using the following:

{\singlespacing
\begin{lstlisting}[basicstyle=\scriptsize\ttfamily]
> install.packages("nprobust")
\end{lstlisting}}

Additionally, the package help manual can be accessed via: 

{\singlespacing
\begin{lstlisting}[basicstyle=\scriptsize\ttfamily]
> help(package = "nprobust")
\end{lstlisting}}

Once the package has been installed, it can be loaded by typing:

{\singlespacing
\begin{lstlisting}[basicstyle=\scriptsize\ttfamily]
> library(nprobust)
\end{lstlisting}}

\citet{Efron-Feldman_1991_JASA} dataset includes five variables: \texttt{t} is an indicator of treatment assignment, \texttt{comp} is a measure of post-treatment compliance, \texttt{chol1} and \texttt{chol2} are two measures of pre-treatment measurements of cholesterol, and \texttt{cholf} records the average of post-treatment measurements of cholesterol. Basic summary statistics on these variables are as follows:

{\singlespacing
\begin{lstlisting}[basicstyle=\scriptsize\ttfamily]
> chole = read.csv("nprobust_data.csv")
> summary(chole)
       t               comp            chol1           chol2           cholf      
 Min.   :0.0000   Min.   :  0.00   Min.   :247.0   Min.   :224.0   Min.   :167.0  
 1st Qu.:0.0000   1st Qu.: 37.00   1st Qu.:277.0   1st Qu.:268.0   1st Qu.:247.0  
 Median :0.0000   Median : 83.00   Median :291.0   Median :281.0   Median :270.0  
 Mean   :0.4896   Mean   : 67.37   Mean   :297.1   Mean   :288.4   Mean   :269.9  
 3rd Qu.:1.0000   3rd Qu.: 96.00   3rd Qu.:306.0   3rd Qu.:298.0   3rd Qu.:288.0  
 Max.   :1.0000   Max.   :101.00   Max.   :442.0   Max.   :435.0   Max.   :427.0  
\end{lstlisting}}

We first employ \code{kdrobust()} with its default options to depict the distribution of the pre-intervention and post-intervention variables. Recall that this command is only valid for interior points, because of the well-known boundary bias in standard density estimation, so the grid of evaluation points needs to be restricted accordingly. As a first illustration, and to display the output in a parsimonious way, we use the function \code{kdrobust()} to estimate the density function of the pre-treatment variable \texttt{chol1} among control units over only seven evaluation points. The command and its summary output is as follows:

{\singlespacing
\begin{lstlisting}[basicstyle=\scriptsize\ttfamily]
> f0_chol1 = kdrobust(chol1, subset = (t == 0), neval = 7)
> summary(f0_chol1)
Call: kdrobust

Sample size (n)                            =     172
Kernel order for point estimation (p)      =     2
Kernel function                            =     Epanechnikov
Bandwidth selection method                 =     imse-dpi

=============================================================================
                                     Point      Std.       Robust B.C.       
          eval        bw   Eff.n      Est.     Error      [ 95% C.I. ]       
=============================================================================
1      269.000    29.291      89     0.011     0.001     [0.010 , 0.014]     
2      277.000    29.291     105     0.014     0.001     [0.013 , 0.018]     
3      284.000    29.291     125     0.016     0.001     [0.015 , 0.020]     
4      291.000    29.291     129     0.016     0.001     [0.016 , 0.020]     
5      300.000    29.291     107     0.014     0.001     [0.013 , 0.018]     
-----------------------------------------------------------------------------
6      306.000    29.291      92     0.012     0.001     [0.010 , 0.015]     
7      329.800    29.291      27     0.004     0.001     [0.001 , 0.004]     
=============================================================================
\end{lstlisting}}

Naturally, as the number of evaluation points increases the output via the \proglang{R} function \code{summary()} increases as well. (In \proglang{Stata} there is an option to suppress output, as in that platform commands typically issue detailed output by default.) Notice that the number of evaluation points was specified, via the option \code{neval = 7}, but not their location. The alternative option \code{eval} controls the number and location of evaluation points: for example, specifying \code{eval = 3} would result in estimation at $x=3$ only or specifying \code{eval = seq(1,2,.1)} would result in estimation at the eleven evaluation points $x\in\{1.0, 1.1, \cdots,1.9,2.0\}$. Unless the option \code{h} is specified, \code{kdrobust()} employs the companion function \code{kdbwselect()}, with its default option \code{bwselect = "IMSE-DPI"}, to choose a data-driven bandwidth; that is, unless the user specifies the bandwidth(s) manually, the default choice is an automatic DPI implementation of the IMSE-optimal bandwidth. 

We do not discuss other details of the output above to conserve space, and mainly because the generic structure and underlying ideas will be discussed in detail further below when employing the function \code{lprobust()}; both commands and outputs have the exact same structure. The key outputs needed at this point are the density point estimates (under \texttt{Point Est.}) and their companion robust bias-corrected confidence intervals (under \texttt{Robust B.C. [95\% C.I.]}), since these ingredients are needed for plotting the estimated density functions.

To now plot and compare the density functions for the different variables and groups, we use the function \code{kdrobust()} multiple times across a larger number of evaluation points, and store the results accordingly:

{\singlespacing
\begin{lstlisting}[basicstyle=\scriptsize\ttfamily]
> ev1 = seq(250, 350, length.out = 30)
> ev2 = seq(0, 100, length.out = 30)
> f0_chol1 = kdrobust(chol1, subset = (t == 0), eval = ev1)
> f1_chol1 = kdrobust(chol1, subset = (t == 1), eval = ev1)
> f0_chol2 = kdrobust(chol2, subset = (t == 0), eval = ev1)
> f1_chol2 = kdrobust(chol2, subset = (t == 1), eval = ev1)
> f0_cholf = kdrobust(cholf, subset = (t == 0), eval = ev1)
> f1_cholf = kdrobust(cholf, subset = (t == 1), eval = ev1)
> f0_comp = kdrobust(comp, subset = (t == 0), eval = ev2)
> f1_comp = kdrobust(comp, subset = (t == 1), eval = ev2)
\end{lstlisting}}

\begin{figure}[!htp]
	\centering
	\subfloat[Pre-treatment: \texttt{chol1}]{\resizebox{0.5\columnwidth}{!}{\includegraphics{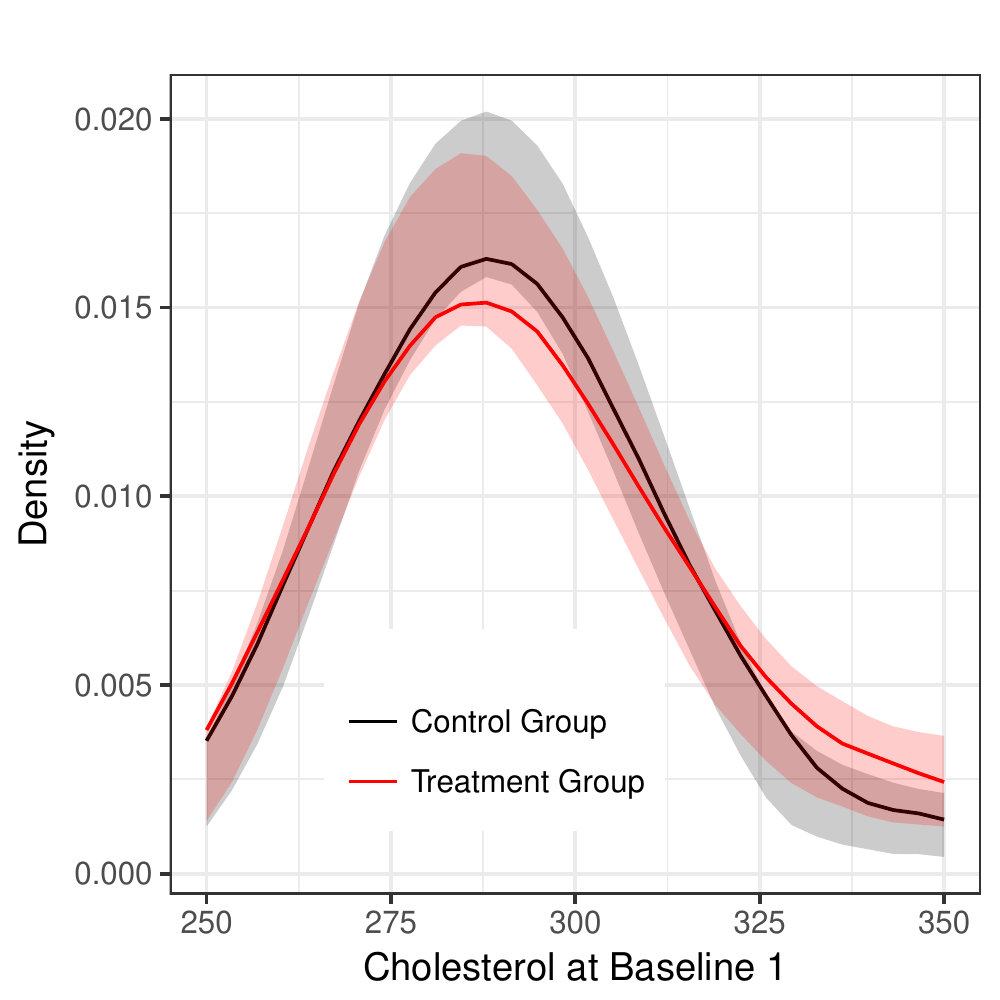}}}
	\subfloat[Pre-treatment: \texttt{chol2}]{\resizebox{0.5\columnwidth}{!}{\includegraphics{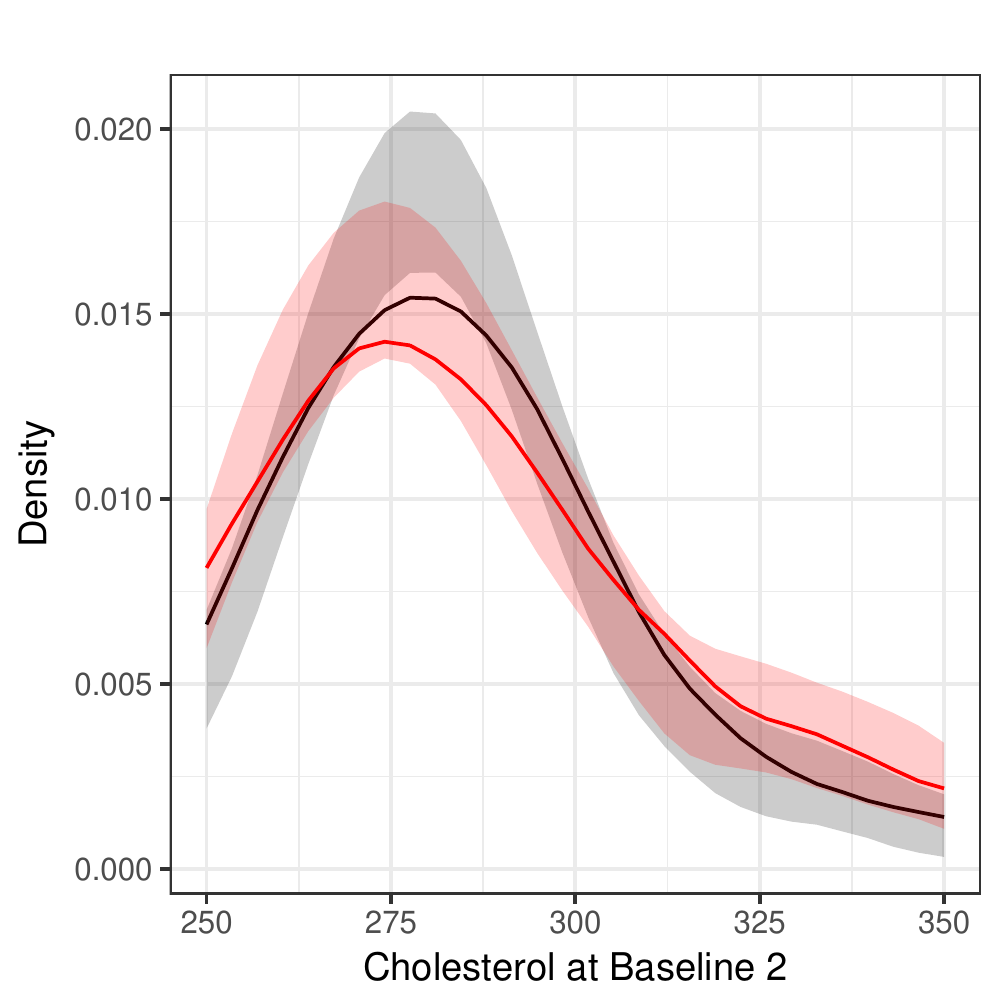}}} \\
	\subfloat[Post-treatment: \texttt{cholf}]{\resizebox{0.5\columnwidth}{!}{\includegraphics{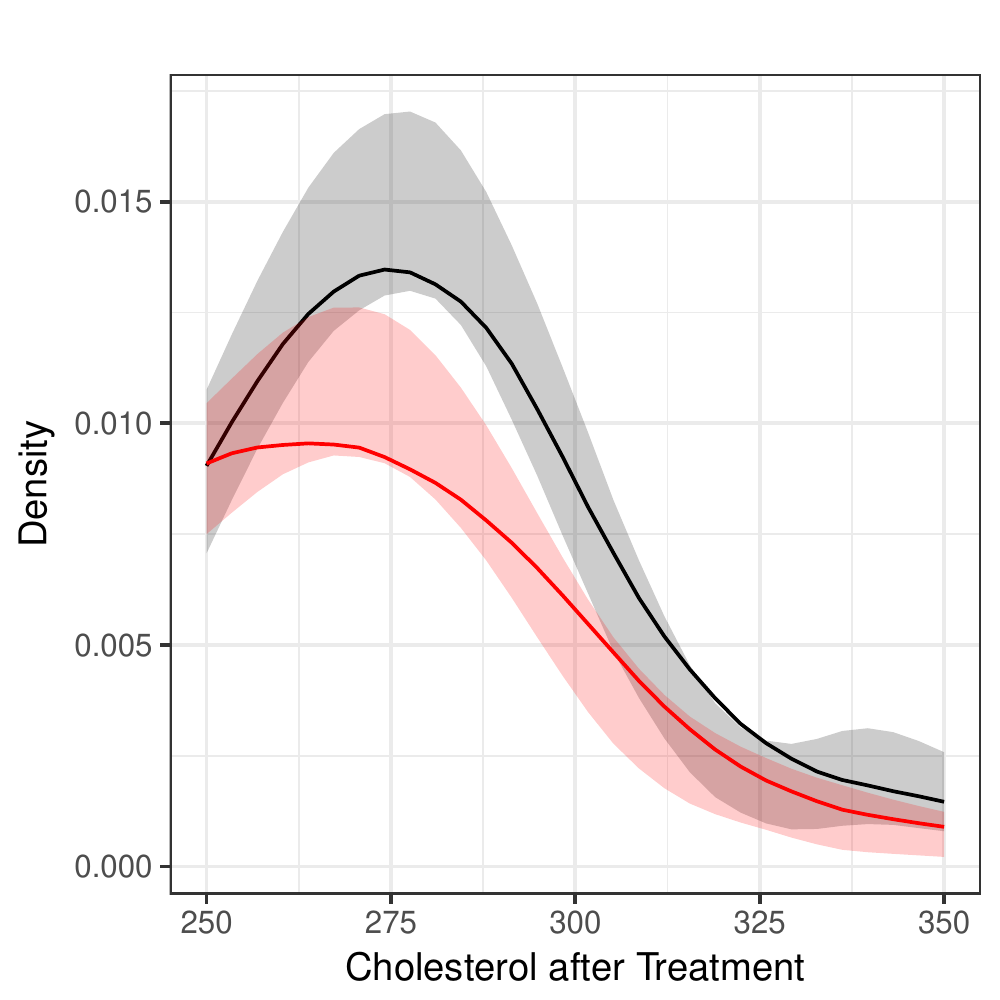}}}
	\subfloat[Post-treatment: \texttt{comp}]{\resizebox{0.5\columnwidth}{!}{\includegraphics{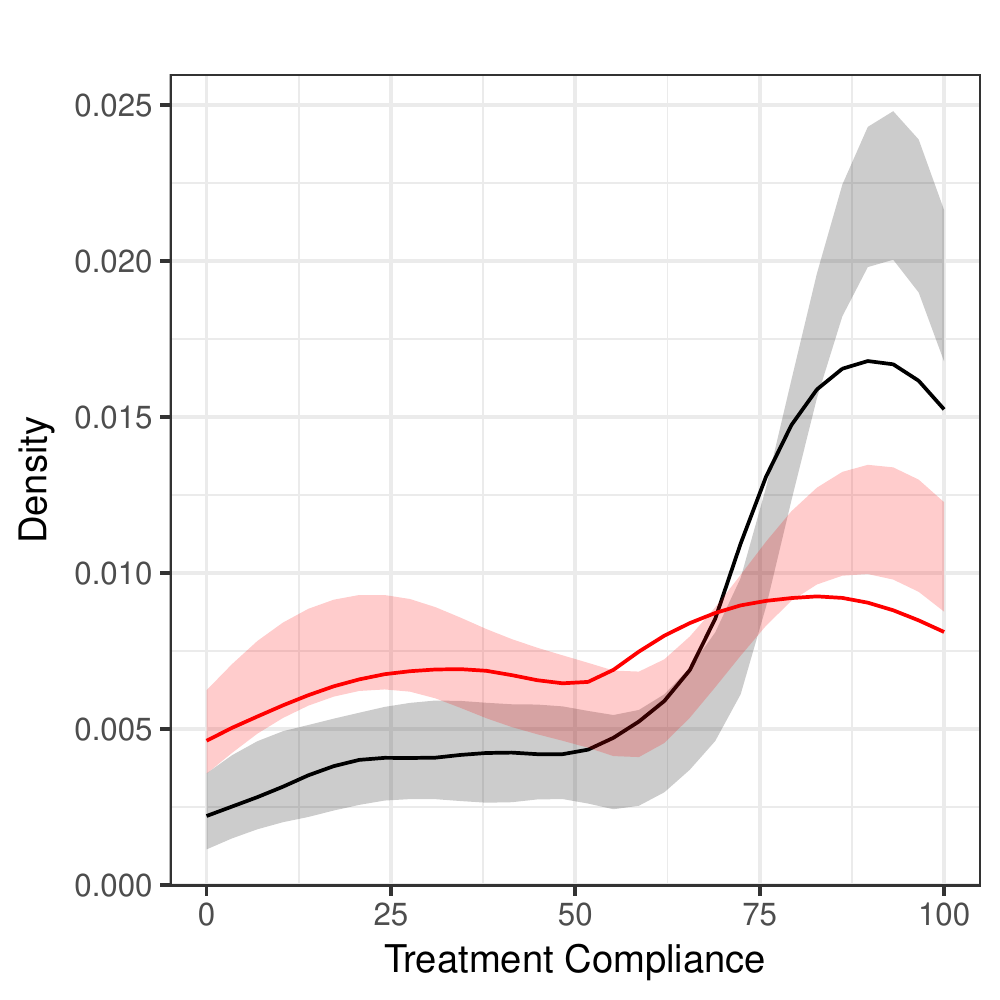}}} \\
	\caption{Kernel Density Estimation with Robust Bias-Corrected Confidence Intervals using IMSE-DPI optimal bandwidth.\label{fig:kdrobust}}
\end{figure}

Given the above information, we can easily plot and compare the estimated densities with the function \code{kdrobust.plot()}, giving the results reported in Figure \ref{fig:kdrobust}:

{\singlespacing
\begin{lstlisting}[basicstyle=\scriptsize\ttfamily]
> nprobust.plot(f0_chol1, f1_chol1, legendGroups = c("Control Group", 
+ "Treatment Group"), xlabel = "Cholesterol at Baseline 1", 
+ ylabel = "Density") + theme(legend.position = c(0.4, 0.2))
> ggsave("output/kd-chol1.pdf", width = 4, height = 4)
> nprobust.plot(f0_chol2, f1_chol2, xlabel = "Cholesterol at Baseline 2", 
+ ylabel = "Density") + theme(legend.position = "none")
> ggsave("output/kd-chol2.pdf", width = 4, height = 4)
> nprobust.plot(f0_cholf, f1_cholf, xlabel = "Cholesterol after Treatment", 
+ ylabel = "Density") + theme(legend.position = "none")
> ggsave("output/kd-cholf.pdf", width = 4, height = 4)
> nprobust.plot(f0_comp, f1_comp, xlabel = "Treatment Compliance", 
+ ylabel = "Density") + theme(legend.position = "none")
> ggsave("output/kd-comp.pdf", width = 4, height = 4)
\end{lstlisting}}

In Figure \ref{fig:kdrobust}, the density functions with red and grey confidence intervals correspond to the control and treatment groups, respectively. Not surprisingly, this figure confirms that \texttt{chol1} and \texttt{chol2} are pre-intervention variables: the have very similar distributions for control and treatment groups. In contrast, both post-treatment cholesterol and compliance exhibit important (and statistically significant) changes between the control and treatment groups.

The average (intention-to-treat) treatment effect is about $-26$ cholesterol points and highly statistically significant. This effect amounts roughly to a $10\%$ reduction in cholesterol given the baseline measurements \texttt{chol1} and \texttt{chol2}. In addition, the average treatment compliance is $-15$ percentage points. However, these are average effects not accounting for observed (treatment effect) heterogeneity across pre-intervention cholesterol levels.

We now illustrate the main functions for local polynomial regression estimation in the package \pkg{nprobust} by analyzing heterogeneous treatment effects. The function \code{lprobust()} provides point estimates and robust confidence intervals employing local polynomial estimators, given a grid of evaluation points and a bandwidth choice. If the evaluation points are not provided, 30 equally-spaced points are chosen over the full support of the data. In this empirical illustration, however, there are a very few observations at the extreme values of the pre-intervention covariates \texttt{chol1} and \texttt{chol2} (see Figure \ref{fig:kdrobust}(a)-(b)), and therefore we restrict the support of analysis to $[250,350]$, which also ensure a common support across both pre-intervention variables. The following commands estimate local polynomial regressions at $30$ evenly-spaced evaluation points over the restricted support using the (default) IMSE-optimal plug-in bandwidth selector computed by \code{lpbwselect()}:

{\singlespacing
\begin{lstlisting}[basicstyle=\scriptsize\ttfamily]
> ev = seq(250, 350, length.out = 30)
> m0_cholf_1 = lprobust(cholf, chol1, subset = (t == 0), eval = ev)
> m1_cholf_1 = lprobust(cholf, chol1, subset = (t == 1), eval = ev)
> m0_comp_1 = lprobust(comp, chol1, subset = (t == 0), eval = ev)
> m1_comp_1 = lprobust(comp, chol1, subset = (t == 1), eval = ev)
> m0_cholf_2 = lprobust(cholf, chol2, subset = (t == 0), eval = ev)
> m1_cholf_2 = lprobust(cholf, chol2, subset = (t == 1), eval = ev)
> m0_comp_2 = lprobust(comp, chol2, subset = (t == 0), eval = ev)
> m1_comp_2 = lprobust(comp, chol2, subset = (t == 1), eval = ev)
\end{lstlisting}}

As before, the results are stored and outputs are not displayed because of their length. They look exactly like the output presented above when using the function \code{kdrobust()}. For instance, using the command \code{summary(m0\_cholf\_1)} after the commands above results in an output including a local polynomial regression estimate of the outcome variable \texttt{cholf} given the pre-intervention covariate \texttt{chol1} for the control group ($\mathtt{t}=0$) over the $30$ evaluation generated by the \proglang{R} function \code{seq(250,350,length.out=30)}. Here is an example of the output, restricted to the first seven evaluation points:

{\singlespacing
\begin{lstlisting}[basicstyle=\scriptsize\ttfamily]
> summary(lprobust(cholf, chol1, subset = (t == 0), eval = ev[1:7]))
Call: lprobust

Sample size (n)                              =    172
Polynomial order for point estimation (p)    =    1
Order of derivative estimated (deriv)        =    0
Polynomial order for confidence interval (q) =    2
Kernel function                              =    Epanechnikov
Bandwidth method                             =    imse-dpi

=============================================================================
                                     Point      Std.       Robust B.C.       
          eval         h   Eff.n      Est.     Error      [ 95% C.I. ]       
=============================================================================
1      250.000   126.950     165   238.530     3.002   [235.157 , 253.518]   
2      253.448   126.950     166   241.657     2.785   [238.453 , 254.069]   
3      256.897   126.950     167   244.856     2.570   [241.732 , 254.944]   
4      260.345   126.950     167   248.122     2.358   [244.988 , 256.170]   
5      263.793   126.950     167   251.386     2.159   [248.320 , 257.860]   
-----------------------------------------------------------------------------
6      267.241   126.950     167   254.650     1.973   [251.672 , 259.886]   
7      270.690   126.950     167   257.913     1.801   [255.008 , 262.183]   
=============================================================================
\end{lstlisting}}

The first panel of the output provides basic information on the options specified in the function. The default estimand is the regression function, indicated by \texttt{Order of derivative estimated (deriv) = 0}. The main panel of the output gives estimation results: (i) \texttt{eval} is the grid of evaluation points; (ii) \texttt{h} is the bandwidth used for point estimation; (iii) \texttt{Eff.n} is the effective sample size (determined by \texttt{h}); (iv) \texttt{Est.} is the point estimate using polynomial order $p$, that is, $\hat{m}^{(0)}(x)$ using the notation above; (iv) \texttt{Std.\ Error} is the standard error of the point estimate, that is, $\hat{\mathscr{V}}[\hat{m}^{(0)}(x)]$ using the notation above; and (v) \texttt{Robust B.C. [95\% C.I.]} is the robust bias-corrected confidence interval, that is, $I_\RBC(x)$ using the notation above.

\begin{figure}[!htp]
	\centering
	\subfloat[Treatment Effects by \texttt{chol1}]{\resizebox{0.5\columnwidth}{!}{\includegraphics{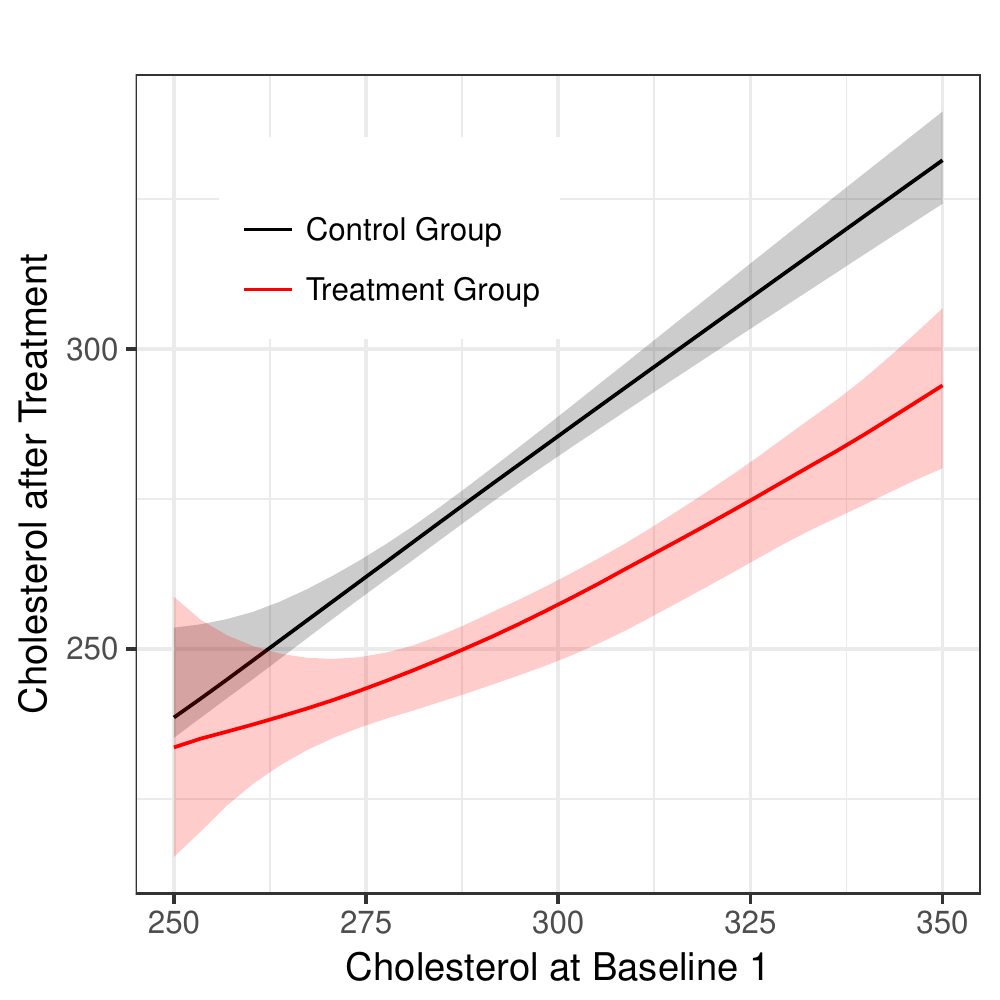}}}
	\subfloat[Treatment Effects by \texttt{chol2}]{\resizebox{0.5\columnwidth}{!}{\includegraphics{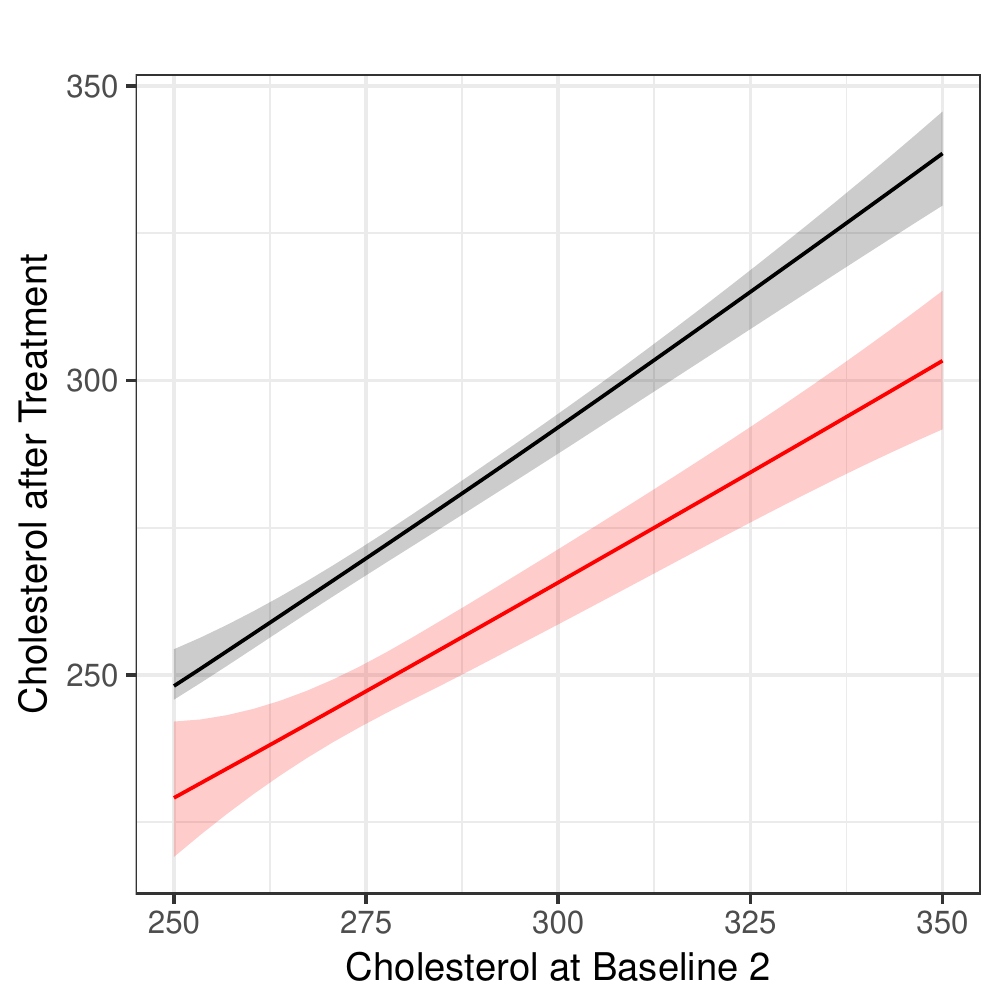}}}\\
	\subfloat[Treatment Complaince by \texttt{chol1}]{\resizebox{0.5\columnwidth}{!}{\includegraphics{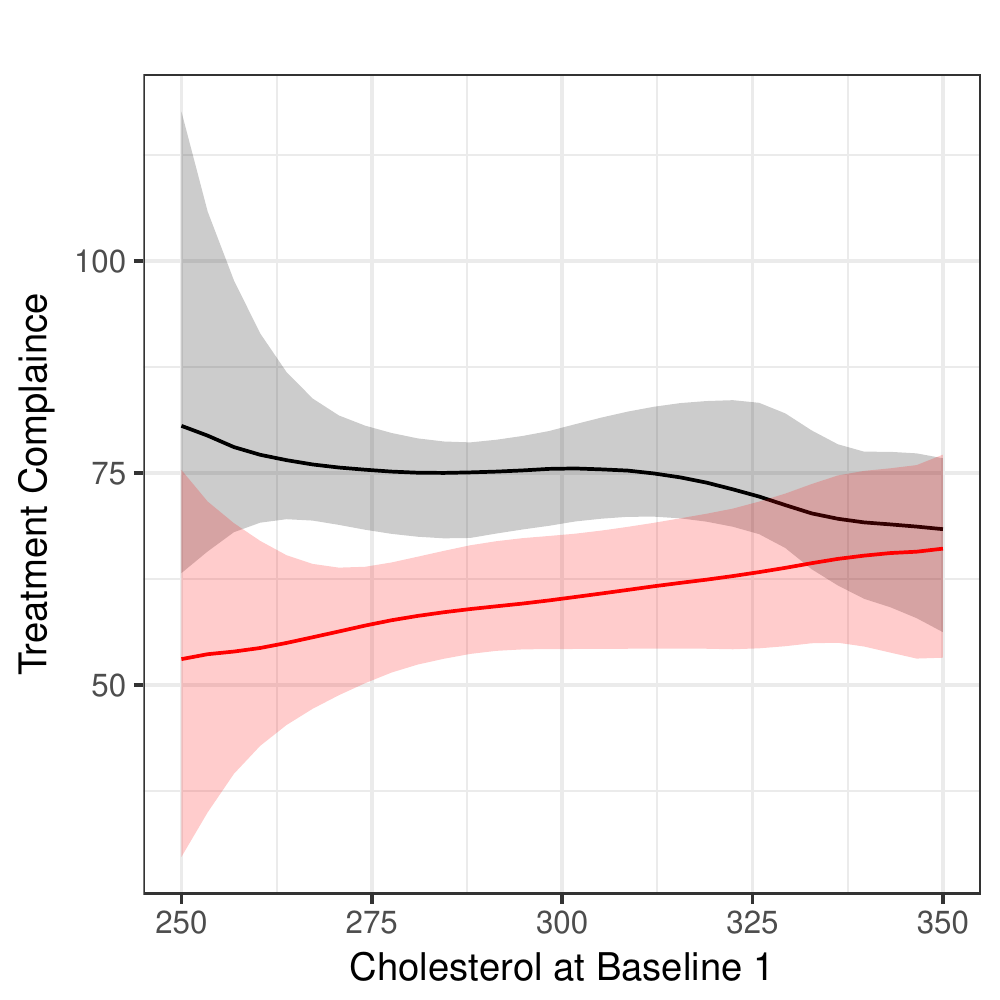}}} 
	\subfloat[Treatment Complaince by \texttt{chol2}]{\resizebox{0.5\columnwidth}{!}{\includegraphics{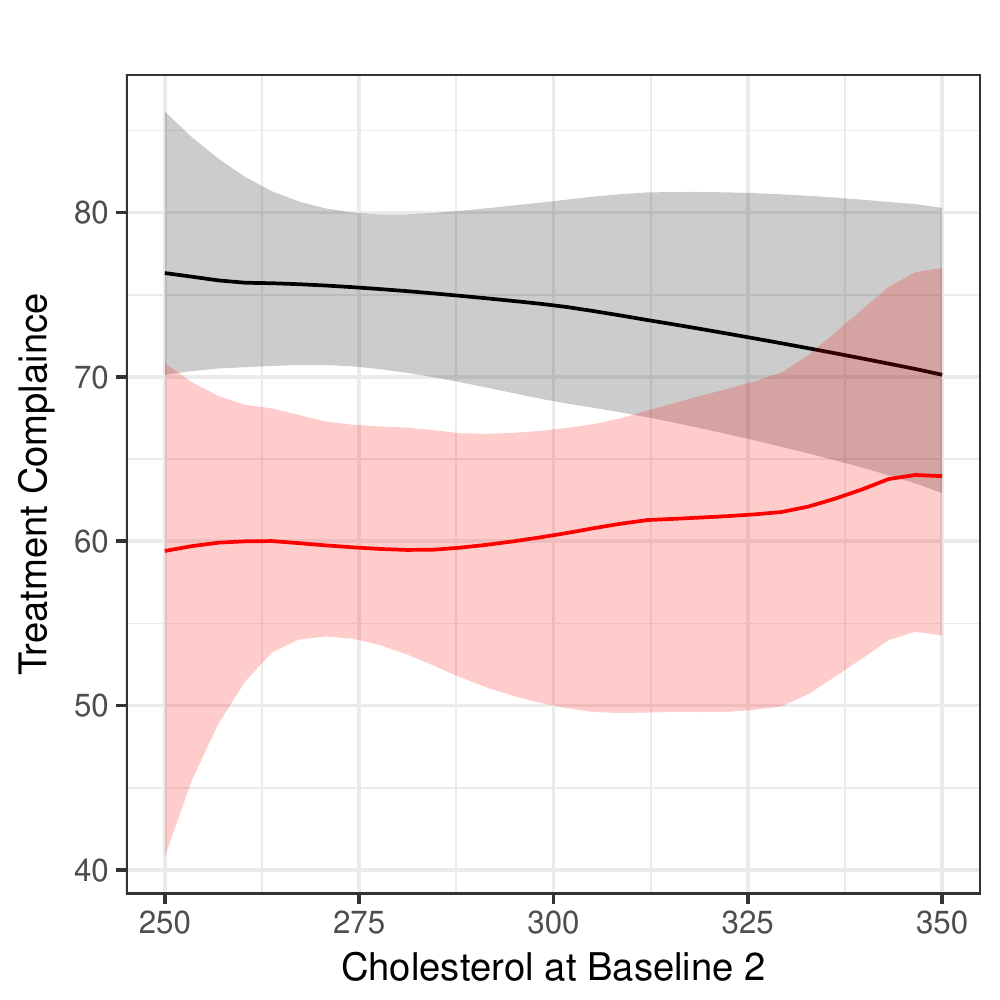}}}\\
	\caption{Local Polynomial Regression with Robust Bias-Corrected Confidence Intervals using IMSE-DPI optimal bandwidth.\label{fig:lprobust}}
\end{figure}

We discuss alternative choices for implementation, estimation, and inference further below, but first we plot the results obtained above using the selected grid of evaluation points. This is with the following commands leading to Figure \ref{fig:lprobust}:

{\singlespacing
\begin{lstlisting}[basicstyle=\scriptsize\ttfamily]
> nprobust.plot(m0_cholf_1, m1_cholf_1, legendGroups = c("Control Group", 
+ "Treatment Group"), xlabel = "Cholesterol at Baseline 1", 
+ ylabel = "Cholesterol after Treatment") + theme(legend.position = c(0.3, 
+ 0.8))
> ggsave("output/lp-cholf-1.pdf", width = 4, height = 4)
> nprobust.plot(m0_cholf_2, m1_cholf_2, xlabel = "Cholesterol at Baseline 2", 
+ ylabel = "Cholesterol after Treatment") + theme(legend.position = "none")
> ggsave("output/lp-cholf-2.pdf", width = 4, height = 4)
> nprobust.plot(m0_comp_1, m1_comp_1, xlabel = "Cholesterol at Baseline 1", 
+ ylabel = "Treatment Complaince") + theme(legend.position = "none")
> ggsave("output/lp-comp-1.pdf", width = 4, height = 4)
> nprobust.plot(m0_comp_2, m1_comp_2, xlabel = "Cholesterol at Baseline 2", 
+ ylabel = "Treatment Complaince") + theme(legend.position = "none")
> ggsave("output/lp-comp-2.pdf", width = 4, height = 4)
\end{lstlisting}}

The results obtained in Figure \ref{fig:lprobust} not only illustrate some of the main features of the \pkg{nprobust} package, but also are substantively interesting. We find that the (intention-to-treat) treatment effect is heterogeneous in the sense that there are no statistical significant effects for observations with low levels of pre-intervention cholesterol, while there are strongly statistically significant effects for units with moderate to high levels of pre-treatment cholesterol (subfigures (a) and (b)). Interestingly, the effects are constant when present. Furthermore, we also find heterogeneous treatment compliance where observations with higher levels of pre-intervention cholesterol tend to comply with treatment while units with low to moderate levels tend not to.

As mentioned previously, the default data-driven bandwidth implementation is IMSE-DPI. However, the package \pkg{nprobust} includes other bandwidth selections, including CE-optimal ones as explained above. These alternative bandwidth choices can be selected directly for estimation and inference via the option \code{bwselect} in the function \code{lprobust()}, or they can all be obtained using the companion bandwidth selection function \code{lpbwselect()}. We illustrate the latter by reporting the output underlying the first seven evaluation points used for local polynomial regression estimation of the outcome variable \texttt{cholf} given the pre-intervention covariate \texttt{chol1} for the control group ($\mathtt{t}=0$) over the $30$ evaluation generated by the \proglang{R} function \code{seq(250,350,length.out=30)}:

{\singlespacing
\begin{lstlisting}[basicstyle=\scriptsize\ttfamily]
> summary(lpbwselect(cholf, chol1, subset = (t == 0), eval = ev[1:7]))
Call: lpbwselect

Sample size (n)                              =    172
Polynomial order for point estimation (p)    =    1
Order of derivative estimated (deriv)        =    0
Polynomial order for confidence interval (q) =    2
Kernel function                              =    Epanechnikov
Bandwidth method                             =    mse-dpi

=======================================
      eval       h        b
=======================================
1    250.000  55.942   70.214
2    253.448  70.164   69.902
3    256.897  43.066   73.278
4    260.345  38.760   76.971
5    263.793  34.577   88.766
---------------------------------------
6    267.241  34.990   70.486
7    270.690  31.096   69.618
=======================================
\end{lstlisting}}

To close this section, we showcase all bandwidth selection methods available. So far all the bandwidths were obtained for the choice MSE-DPI or IMSE-DPI. The following output exhibits all possible bandwidths, including the recently introduced CE-optimal bandwidth choices using a DPI selector (CE-DPI).

{\singlespacing
\begin{lstlisting}[basicstyle=\scriptsize\ttfamily]
> summary(lpbwselect(cholf, chol1, subset = (t == 0), eval = ev[1:7], 
+ bwselect = "ALL"))
Call: lpbwselect

Sample size (n)                              =    172
Polynomial order for point estimation (p)    =    1
Order of derivative estimated (deriv)        =    0
Polynomial order for confidence interval (q) =    2
Kernel function                              =    Epanechnikov
Bandwidth method                             =    all

=======================================================================================
                   MSE-DPI           MSE-ROT            CE-DPI            CE-ROT
      eval       h        b        h        b        h        b        h        b
=======================================================================================
1    250.000  55.942   70.214  127.743   20.375   63.010   44.433   43.248   44.433
2    253.448  70.164   69.902  111.187   20.478   87.011   44.236   54.243   44.236
3    256.897  43.066   73.278   82.630   20.745   97.192   46.373   33.293   46.373
4    260.345  38.760   76.971   69.839   20.519   69.521   48.709   29.965   48.709
5    263.793  34.577   88.766   65.155   20.856   61.632   56.174   26.731   56.174
---------------------------------------------------------------------------------------
6    267.241  34.990   70.486   61.440   21.044   59.552   44.606   27.050   44.606
7    270.690  31.096   69.618   59.827   21.471   53.157   44.057   24.040   44.057
=======================================================================================

===================================
         IMSE-DPI          IMSE-ROT
       h        b        h        b
===================================
  94.694   87.874   77.135   29.077
===================================

\end{lstlisting}}

This output gives MSE, IMSE, and CE bandwidth selectors, implemented using either DPI or a ROT approach. The above bandwidth selection execution included the (default) option \code{bwcheck=21}. This option forces the command to select the minimum bandwidth such that at least $21$ observations are used at each evaluation point, which is useful to mitigate issues related to sparse data at certain evaluation points. We combined this option with the IMSE-DPI bandwidth because it employs a larger grid of evaluation points to approximate the outer integrals present in its construction, and some of these grid points exhibited issues of (local) sparsity in this particular application.

In the upcoming section we illustrate how these bandwidth selection methods, together with the estimation and inference procedures discussed above, perform in simulations. We also show how they compare to other available \proglang{R} packages.

\section[Simulations]{Simulations and Comparisons with Other \proglang{R} Packages}\label{sec:simuls}

In this section we present the results from a simulation study addressing the finite-sample performance of \pkg{nprobust} in comparison to other \proglang{R} packages implementing kernel smoothing regression methods. We consider four other packages: \pkg{locfit} \citep{locfit}, \pkg{locpol}, \pkg{LPridge}, and \pkg{np} \citep{npJSS}; Table \ref{table:Rpkgs} gives a brief comparison of the main features available in each of the packages. Among the main differences summarized there, the most important one is related to inference: none of the other packages account for the first-order asymptotic bias present in the distributional approximation when employing automatic MSE-optimal or similar bandwidth selection. In addition, none of the other packages employ pre-asymptotic quantities when constructing the estimators and test statistics. As we show below, these differences in implementation will lead to substantial empirical coverage distortions of confidence intervals when contrasted with the performance of \pkg{nprobust}. See the Appendix for a comparison of \proglang{R} syntax across these packages.

We compare the performance in finite samples of the five packages described in Table \ref{table:Rpkgs} using a simulation study. The data generating process was previously used by \citet{Fan-Gijbels_1996_Book} and \citet{Cattaneo-Farrell2013_JoE}, and is given as follows:
\[Y = m(x)+\varepsilon,\qquad
m(x) = \sin(2x-1) + 2\exp\{-16(x-0.5)^2\}\qquad
x\thicksim\mathcal{U}[0,1], \qquad
\varepsilon\thicksim\mathcal{N}(0,1).
\]
We consider $5,000$ simulations, where for each replication the data is generated as i.i.d. draws of size $n=500$. We investigate several performance measures for point estimation and inference, whenever available in each package. Specifically, we consider bias, variance, MSE, and empirical coverage and average interval length of nominal $95\%$ confidence intervals, over five equally-spaced values $x\in\{0,0.25,0.5,0.75,1\}$. The regression function and evaluation points considered are plotted in Figure \ref{fig:regfun}, together with a realization of the data and regression function estimate (and confidence intervals) using \pkg{nprobust}. Whenever possible we employ the Epanechnikov kernel, but otherwise whatever is the default kernel in each package.

We focus on two types of comparisons across packages in terms of point estimation and inference performance: (i) using a fixed, MSE-optimal bandwidth choice, and (ii) using an estimated bandwidth. This allows to disentangle the effects of constructing point estimators and inference procedures (given a fixed, common choice of bandwidth across packages) from the convoluted effects of different point estimators and inference procedures together with different data-driven bandwidth selectors.

We organize the presentation of the results by choice of polynomial order ($p$) and derivative being estimated ($\nu$) in five tables as follows:
\begin{itemize}
	\singlespacing
	\item Table \ref{table:simuls_p1d0}: local linear estimator ($p=1$) for the level of the regression function ($\nu=0$);
	\item Table \ref{table:simuls_p0d0}: local constant estimator ($p=0$) for the level of the regression function ($\nu=0$);
	\item Table \ref{table:simuls_p2d0}: local quadratic estimator ($p=2$) for the level of the regression function ($\nu=0$);
	\item Table \ref{table:simuls_p1d1}: local linear estimator ($p=1$) for the derivative of the regression function ($\nu=1$);
	\item Table \ref{table:simuls_p2d1}: local quadratic estimator ($p=2$) for the derivative of the regression function ($\nu=1$).
\end{itemize}

Recall that $p=1$ and $\nu=0$ is the default for the package \pkg{nprobust}, and corresponds to the recommended $p-\nu$ odd case for estimating $m(x)$; similarly, Table \ref{table:simuls_p2d1} presents the results for the recommended $p-\nu$ odd case for estimating $m^{(1)}(x)$. Each of these tables reports results on the performance of each of the four packages available in \proglang{R} in addition to the package \pkg{nprobust}, for each of the five evaluation points mentioned previously. For the case of package \pkg{nprobust}, we consider all three bandwidth methods: MSE-optimal, IMSE-optimal, and CE-optimal. All this information is organized by rows in each of the tables.

Each of the Tables \ref{table:simuls_p1d0}--\ref{table:simuls_p2d1} have two sets of columns labeled ``Population Bandwidth'' and ``Estimated Bandwidth'', which correspond to simulation results using either a common MSE-optimal population bandwidth or a different estimated bandwidth as computed by each package. For each of the two groups of collumns, we report: (i) population or average estimated bandwidths under the label ``$h$''; (ii) the average simulation bias of the point estimator under the label ``Bias''; (iii) the average simulation variance of the point estimator under the label ``Var''; (iv) the average simulation MSE of the point estimator under the label ``MSE''; (v) the average simulation empirical coverage of $95\%$ confidence intervals under the label ``EC''; and (vi) the average simulation empirical interval length of $95\%$ confidence intervals under the label ``IL''.

For brevity, here we discuss only the default case $p=1$ and $\nu=0$ (Table \ref{table:simuls_p1d0}), but the results are qualitatively consistent across all tables. We found that \pkg{nprobust} offers a point estimator with similar MSE properties as other packages available, but much better inference performance. In most cases, the empirical coverage of $95\%$ confidence intervals are about correct for the package \pkg{nprobust} (across evaluation points and bandwidth selection methods), while other packages exhibit important coverage distortions. For example, for the evaluation point $x=0$, which exhibits substantial local curvature in the regression function, the package \pkg{np} delivers nominal $95\%$ confidence intervals with an empirical coverage of $0\%$ when the population MSE-optimal bandwidth is used and of $76.2\%$ with an estimated (by cross-validation) bandwidth. In contrast, the empirical coverage when using the package \pkg{nprobust} is about $94\%$ in the same setting, and using any of the population or estimated bandwidth procedures available in this package. These empirical findings are in line with the underlying theory and illustrate the advantages of employing robust bias correction methods for inference \citep{Calonico-Cattaneo-Farrell_2018_JASA}. 

The other Tables \ref{table:simuls_p0d0}--\ref{table:simuls_p2d1} show qualitatively similar empirical findings. Finally, to offer a more complete set of results we include Table \ref{table:simuls_bw}, which reports the performance of all the bandwidth selectors available in the package \pkg{nprobust}. Some of these results were already reported in the previous tables, but Table \ref{table:simuls_bw} allows for an easier comparison across models and methods, in addition to including the performance of the ROT bandwidth selectors (not reported previously). The simulation results show that these bandwidth selectors offer good performance on average relative to their corresponding population target.

\section[Conclusion]{Conclusion}\label{sec:conclusion}

We gave an introduction to the software package \pkg{nprobust}, which offers an array of estimation and inference procedures for nonparametric kernel-based density and local polynomial methods. This package is available in both \proglang{R} and \proglang{Stata} statistical platforms, and further installation details and replication scripts can be found at \url{https://sites.google.com/site/nprobust/}. We also offered a comprehensive simulation study comparing the finite sample performance of the package vis-\`a-vis other packages available in \proglang{R} for kernel-based nonparametric estimation and inference. In particular, we found that the package \pkg{nprobust} offers on par point estimation methods with superior performance in terms of inference.

\section[Acknowledgments]{Acknowledgments}

We thank David Drukker, Yingjie Feng, Guido Imbens, Michael Jansson, Xinwei Ma, Jeffrey Racine, Rocio Titiunik, Gonzalo Vazquez-Bare, and two anonymous reviewers for thoughtful comments and suggestions. Cattaneo gratefully acknowledges financial support from the National Science Foundation through grant SES-1459931. Farrell gratefully acknowledges financial support from the Richard N. Rosett and John E. Jeuck Fellowships.

\begin{appendices}
	
	\section[Syntax Comparison across R Packages]{Syntax Comparison across \proglang{R} Packages}
	
	For illustration and comparison purposes, we present the {\sf R} syntax for all the packages considered to yield approximately the same point estimates. Data setup:\vspace{-.2in}
	{\singlespacing
\begin{lstlisting}[basicstyle=\scriptsize\ttfamily]
> n <- 500
> m.fun <- function(x) {
+ sin(2 * x - 1) + 2 * exp(-16 * (x - 0.5)^2)
+ }
> x <- runif(n)
> u <- rnorm(n)
> y <- m.fun(x) + u
> eval = 0.75
> h = 0.27
> paste("Point Estimate:", m.fun(eval))
[1] "Point Estimate: 1.21518442094709"
\end{lstlisting}}
	\noindent Syntax for \pkg{locfit}:\vspace{-.2in}
	{\singlespacing
\begin{lstlisting}[basicstyle=\scriptsize\ttfamily]
> locfit.reg = locfit(y ~ lp(x, deg = 1, h = h))
> locfit.est = predict(locfit.reg, c(eval), se.fit = TRUE)
> paste("Point Estimate:", locfit.est$fit)
[1] "Point Estimate: 1.18843838932338"
\end{lstlisting}}
	\noindent Syntax for \pkg{locpol}:\vspace{-.2in}
	{\singlespacing
\begin{lstlisting}[basicstyle=\scriptsize\ttfamily]
> lpol.est = locpol(y ~ x, data = data.frame(y, x), bw = h, kernel = EpaK, 
+ deg = 1, xeval = eval)
> paste("Point Estimate:", as.numeric(lpol.est$lpFit[2]))
[1] "Point Estimate: 1.22282088483116"
\end{lstlisting}}
	\noindent Syntax for \pkg{lpridge}:\vspace{-.2in}
	{\singlespacing
\begin{lstlisting}[basicstyle=\scriptsize\ttfamily]
> lpridge.est = lpridge(x, y, bandwidth = h, x.out = eval, order = 1, 
+ ridge = NULL, weight = "epa", mnew = 100, var = TRUE)
> paste("Point Estimate:", lpridge.est$est)
[1] "Point Estimate: 1.22282121618125"
\end{lstlisting}}
	\noindent Syntax for \pkg{np}:\vspace{-.2in}
	{\singlespacing
\begin{lstlisting}[basicstyle=\scriptsize\ttfamily]
> np.est <- npreg(y ~ x, regtype = "ll", bws = h, exdat = eval, 
+ ckertype = "epanechnikov")
> paste("Point Estimate:", np.est$mean)
[1] "Point Estimate: 1.21647949074877"
\end{lstlisting}}
	\noindent Syntax for \pkg{nprobust}:\vspace{-.2in}
	{\singlespacing
\begin{lstlisting}[basicstyle=\scriptsize\ttfamily]
> lp.est = lprobust(y = y, x = x, p = 1, eval = eval, h = h, kernel = "epa")
> paste("Point Estimate:", lp.est$Estimate[, "tau.us"])
[1] "Point Estimate: 1.22282088483116"
\end{lstlisting}}
	
\end{appendices}

\bibliography{Calonico-Cattaneo-Farrell_2019_JSS}

\begin{thebibliography}{19}
\newcommand{\enquote}[1]{``#1''}
\providecommand{\natexlab}[1]{#1}
\providecommand{\url}[1]{\texttt{#1}}
\providecommand{\urlprefix}{URL }
\expandafter\ifx\csname urlstyle\endcsname\relax
  \providecommand{\doi}[1]{doi:\discretionary{}{}{}#1}\else
  \providecommand{\doi}{doi:\discretionary{}{}{}\begingroup
  \urlstyle{rm}\Url}\fi
\providecommand{\eprint}[2][]{\url{#2}}

\bibitem[{Abadie and Imbens(2008)}]{Abadie-Imbens_2008_AdES}
Abadie A, Imbens GW (2008).
\newblock \enquote{Estimation of the Conditional Variance in Paired
  Experiments.}
\newblock \emph{Annales d'Economie et de Statistique}, (91/92), 175--187.

\bibitem[{Calonico \emph{et~al.}(2018)Calonico, Cattaneo, and
  Farrell}]{Calonico-Cattaneo-Farrell_2018_JASA}
Calonico S, Cattaneo MD, Farrell MH (2018).
\newblock \enquote{On the Effect of Bias Estimation on Coverage Accuracy in
  Nonparametric Inference.}
\newblock \emph{Journal of the American Statistical Association},
  \textbf{113}(522), 767--779.

\bibitem[{Calonico \emph{et~al.}(2019)Calonico, Cattaneo, and
  Farrell}]{Calonico-Cattaneo-Farrell_2019_wp}
Calonico S, Cattaneo MD, Farrell MH (2019).
\newblock \enquote{Coverage Error Optimal Confidence Intervals for Local
  Polynomial Regression.}
\newblock \emph{\emph{arXiv:1808.01398}}.

\bibitem[{Cattaneo \emph{et~al.}(2019{\natexlab{a}})Cattaneo, Crump, Farrell,
  and Feng}]{Cattaneo-Crump-Farrell-Feng_2019_Stata}
Cattaneo MD, Crump RK, Farrell MH, Feng Y (2019{\natexlab{a}}).
\newblock \enquote{Binscatter Regressions.}
\newblock \emph{\emph{arXiv:1902.09615}}.

\bibitem[{Cattaneo and Farrell(2013)}]{Cattaneo-Farrell2013_JoE}
Cattaneo MD, Farrell MH (2013).
\newblock \enquote{Optimal Convergence Rates, Bahadur Representation, and
  Asymptotic Normality of Partitioning Estimators.}
\newblock \emph{Journal of Econometrics}, \textbf{174}, 127--143.

\bibitem[{Cattaneo \emph{et~al.}(2019{\natexlab{b}})Cattaneo, Farrell, and
  Feng}]{Cattaneo-Farrell-Feng_2019_lspartition}
Cattaneo MD, Farrell MH, Feng Y (2019{\natexlab{b}}).
\newblock \enquote{\texttt{lspartition}: Partitioning-Based Least Squares
  Regression.}
\newblock \emph{\emph{Working paper}}.

\bibitem[{Cattaneo \emph{et~al.}(2019{\natexlab{c}})Cattaneo, Jansson, and
  Ma}]{Cattaneo-Jansson-Ma_2019_lpdensity}
Cattaneo MD, Jansson M, Ma X (2019{\natexlab{c}}).
\newblock \enquote{\texttt{lpdensity}: Local Polynomial Density Estimation and
  Inference.}
\newblock \emph{\emph{Working paper}}.

\bibitem[{Eddelbuettel and Balamuta(2017)}]{Rcpp}
Eddelbuettel D, Balamuta JJ (2017).
\newblock \enquote{{Extending \texttt{R} with \texttt{C++}: A Brief
  Introduction to \texttt{Rcpp}}.}
\newblock \emph{PeerJ Preprints}, \textbf{5}, e3188v1.
\newblock ISSN 2167-9843.
\newblock \doi{10.7287/peerj.preprints.3188v1}.
\newblock \urlprefix\url{https://doi.org/10.7287/peerj.preprints.3188v1}.

\bibitem[{Efron and Feldman(1991)}]{Efron-Feldman_1991_JASA}
Efron B, Feldman D (1991).
\newblock \enquote{Compliance as an Explanatory Variable in Clinical Trials.}
\newblock \emph{Journal of the American Statistical Association},
  \textbf{86}(413), 9--17.

\bibitem[{Fan and Gijbels(1996)}]{Fan-Gijbels_1996_Book}
Fan J, Gijbels I (1996).
\newblock \emph{Local Polynomial Modelling and Its Applications}.
\newblock Chapman \& Hall/CRC, New York.

\bibitem[{Hayfield and Racine(2008)}]{npJSS}
Hayfield T, Racine J (2008).
\newblock \enquote{Nonparametric Econometrics: The np Package.}
\newblock \emph{Journal of Statistical Software}, \textbf{27}(1), 1--32.

\bibitem[{Li and Racine(2007)}]{Li-Racine_2007_Book}
Li Q, Racine S (2007).
\newblock \emph{Nonparametric Econometrics}.
\newblock Princeton University Press, New Yersey.

\bibitem[{Loader(2013)}]{locfit}
Loader C (2013).
\newblock \emph{locfit: Local Regression, Likelihood and Density Estimation.}
\newblock R package version 1.5-9.1,
  \urlprefix\url{http://CRAN.R-project.org/package=locfit}.

\bibitem[{Long and Ervin(2000)}]{Long-Ervin_2000_AS}
Long JS, Ervin LH (2000).
\newblock \enquote{Using Heteroscedasticity Consistent Standard Errors in the
  Linear Regression Model.}
\newblock \emph{The American Statistician}, \textbf{54}(3), 217--224.

\bibitem[{MacKinnon(2012)}]{MacKinnon_2012_BookCh}
MacKinnon JG (2012).
\newblock \enquote{Thirty years of heteroskedasticity-robust inference.}
\newblock In X~Chen, NR~Swanson (eds.), \emph{Recent Advances and Future
  Directions in Causality, Prediction, and Specification Analysis}. Springer.

\bibitem[{Muller and Stadtmuller(1987)}]{Muller-Stadtmuller_1987_AoS}
Muller HG, Stadtmuller U (1987).
\newblock \enquote{Estimation of Heteroscedasticity in Regression Analysis.}
\newblock \emph{Annals of Statistics}, \textbf{15}(2), 610--625.

\bibitem[{Ruppert \emph{et~al.}(2009)Ruppert, Wand, and
  Carroll}]{Ruppert-Wand-Carroll_2009_book}
Ruppert D, Wand MP, Carroll R (2009).
\newblock \emph{Semiparametric Regression}.
\newblock Cambridge University Press, New York.

\bibitem[{Wand and Jones(1995)}]{Wand-Jones_1995_Book}
Wand M, Jones M (1995).
\newblock \emph{Kernel Smoothing}.
\newblock Chapman \& Hall/CRC, Florida.

\bibitem[{Wickham(2016)}]{ggplot2}
Wickham H (2016).
\newblock \emph{ggplot2: Elegant Graphics for Data Analysis}.
\newblock Springer-Verlag New York.
\newblock ISBN 978-3-319-24277-4.
\newblock \urlprefix\url{https://ggplot2.tidyverse.org}.

\end{thebibliography}

\clearpage

\begin{table}
	\caption{Comparison of \proglang{R} Packages Features\label{table:Rpkgs}}\vspace{-.2in}
	\begin{center}\renewcommand{\arraystretch}{1.2}
		\resizebox{\columnwidth}{!}{
			\begin{tabular}{c c c c c c c} 
				\hline
				Package           & Last       & Polynomial  & Derivative    & BW Selection & Standard & Statistical \\
				Name              & Update     & Order $(p)$ & Order $(\nu)$ & Method ($h$) & Errors   & Inference   \\
				\hline\hline
				\pkg{locpol}   & 2018-05-24 & $>0$        & NA            & NA           & NA       & NA \\
				\hline
				\pkg{lpridge}  & 2018-06-12 & $\geq0$     & $\geq0$       & NA           & NA       & NA \\
				\hline
				\pkg{locfit}   & 2013-04-20 & $>0$        & NA            & NN           & HOM      & US \\ 
				\hline
				\pkg{np}       & 2018-10-25 & $0,1$       & $1$           & CV           & HOM      & US \\
				\hline
				\pkg{nprobust} &            & $\geq0$     & $\geq0$       & IMSE/MSE/CE  & HET      & US/RBC \\
				\hline
			\end{tabular}
		}
	\end{center}
	\textit{Notes}: (i) NA means that either the feature is not available or, at least, not documented; (ii) NN means Nearest Neighbor based estimation; (iii) CV means cross-validation based; (iv) HOM means homoskedastic consistent standard errors; (v) HET means heteroskedastic consistent standard errors; (vi) US means undersmoothing inference (not valid when using CV/MSE/IMSE type bandwidth selection); and (vii) RBC means robust bias corrected inference (valid when using any MSE optimization type bandwidth selection). See the Appendix for a comparison of \proglang{R} syntax across these packages.
\end{table}

\begin{figure}
	\centering
	\subfloat[Regression Function and Evaluation Points]{\resizebox{0.5\columnwidth}{3in}{\includegraphics{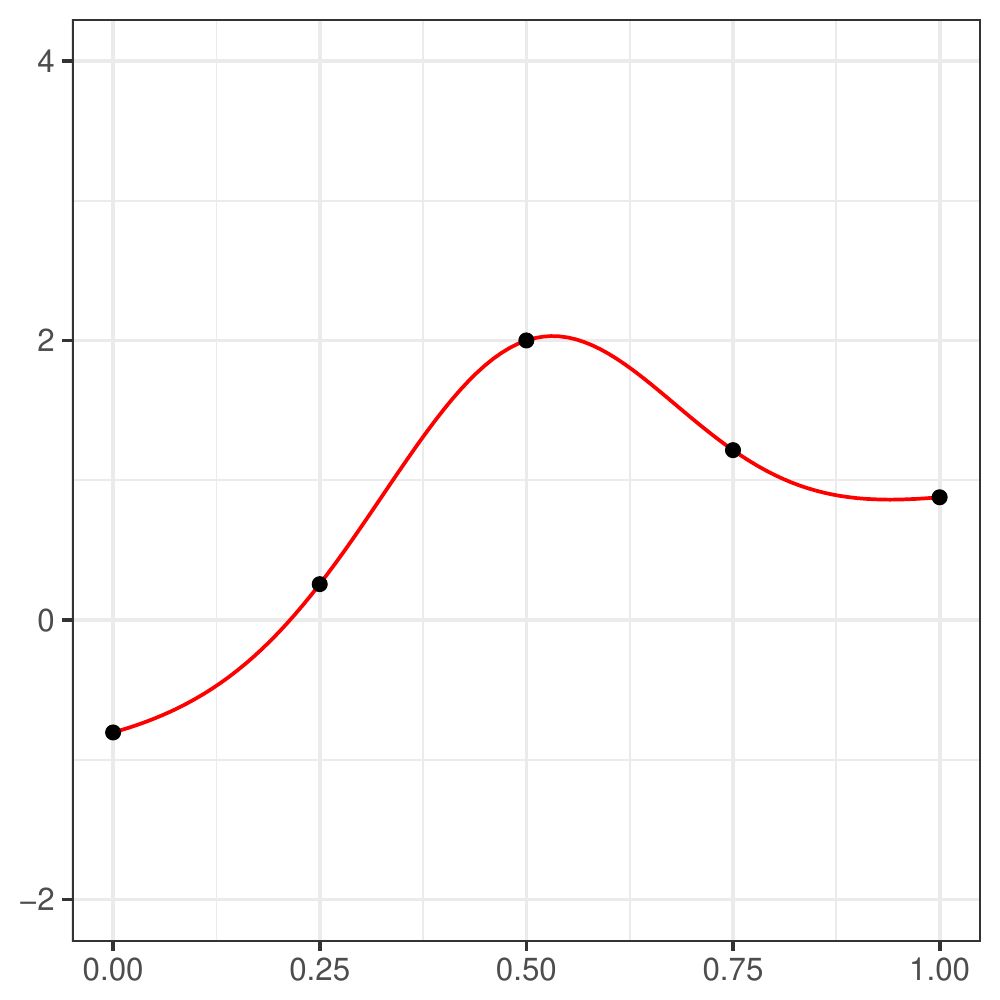}}}
	\subfloat[One Draw of Simulated Data]{\resizebox{0.5\columnwidth}{3.18in}{\includegraphics{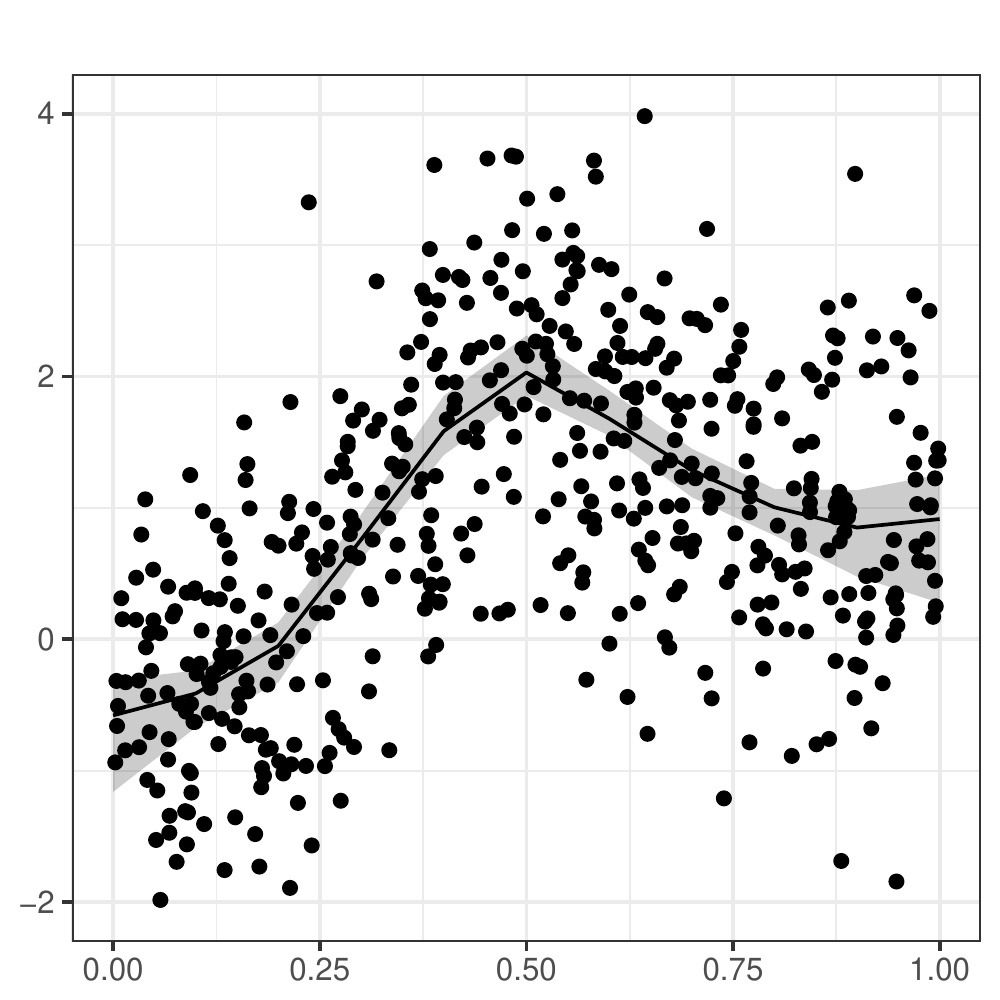}}}
	\caption{Regression Function for Simulations\label{fig:regfun}}\vspace{-.8in}
\end{figure}

\clearpage

\begin{table}\begin{center}\renewcommand{\arraystretch}{1}
		\caption{Simulation Results, $p=1,deriv=0$\label{table:simuls_p1d0}}\vspace{-.1in}
		\resizebox{\columnwidth}{!}{
\begin{tabular}{lccccccccccccc}
\hline\hline
\multicolumn{1}{l}{\bfseries }&\multicolumn{6}{c}{\bfseries Population bandwidth}&\multicolumn{1}{c}{\bfseries }&\multicolumn{6}{c}{\bfseries Estimated bandwidth}\tabularnewline
\cline{2-7} \cline{9-14}
\multicolumn{1}{l}{}&\multicolumn{1}{c}{$h$}&\multicolumn{1}{c}{Bias}&\multicolumn{1}{c}{Var}&\multicolumn{1}{c}{MSE}&\multicolumn{1}{c}{EC}&\multicolumn{1}{c}{IL}&\multicolumn{1}{c}{}&\multicolumn{1}{c}{$h$}&\multicolumn{1}{c}{Bias}&\multicolumn{1}{c}{Var}&\multicolumn{1}{c}{MSE}&\multicolumn{1}{c}{EC}&\multicolumn{1}{c}{IL}\tabularnewline
\hline
{\bfseries $x=$ 0}&&&&&&&&&&&&&\tabularnewline
~~\texttt{locpol}&0.347&0.164&0.026&0.053&1.000&3.917&&0.130&0.018&0.085&0.085&1.000&3.830\tabularnewline
~~\texttt{lpridge}&0.347&0.112&0.023&0.036&0.887&0.601&&\texttt{na}&\texttt{na}&\texttt{na}&\texttt{na}&\texttt{na}&\texttt{na}\tabularnewline
~~\texttt{locfit}&0.347&0.120&0.030&0.044&0.897&0.692&&\texttt{na}&0.312&0.015&0.113&0.293&0.488\tabularnewline
~~\texttt{np}&0.347&0.115&0.012&0.025&0.635&0.309&&0.057&0.025&0.079&0.079&0.692&0.564\tabularnewline
~~\texttt{nprobust}&&&&&&&&&&&&&\tabularnewline
~~      $h_{\texttt{MSE}}$&0.347&0.164&0.026&0.053&0.938&0.928&&0.298&0.035&0.060&0.061&0.904&1.074\tabularnewline
~~     $h_{\texttt{IMSE}}$&0.234&0.063&0.039&0.043&0.929&1.131&&0.182&0.030&0.053&0.054&0.925&1.281\tabularnewline
~~     $h_{\texttt{CE}}$&0.254&0.078&0.036&0.042&0.929&1.084&&0.135&0.019&0.079&0.080&0.910&1.520\tabularnewline
\hline
{\bfseries $x=$ 0.25}&&&&&&&&&&&&&\tabularnewline
~~\texttt{locpol}&0.253&0.116&0.005&0.018&1.000&3.938&&0.130&0.040&0.012&0.013&1.000&3.875\tabularnewline
~~\texttt{lpridge}&0.253&0.116&0.005&0.018&0.610&0.271&&\texttt{na}&\texttt{na}&\texttt{na}&\texttt{na}&\texttt{na}&\texttt{na}\tabularnewline
~~\texttt{locfit}&0.253&0.091&0.006&0.014&0.773&0.296&&\texttt{na}&0.130&0.004&0.021&0.457&0.246\tabularnewline
~~\texttt{np}&0.253&0.096&0.004&0.013&0.770&0.283&&0.057&0.037&0.013&0.014&0.934&0.433\tabularnewline
~~\texttt{nprobust}&&&&&&&&&&&&&\tabularnewline
~~      $h_{\texttt{MSE}}$&0.253&0.116&0.005&0.018&0.942&0.389&&0.163&0.059&0.007&0.011&0.938&0.488\tabularnewline
~~     $h_{\texttt{IMSE}}$&0.234&0.104&0.005&0.016&0.946&0.405&&0.182&0.071&0.007&0.012&0.944&0.459\tabularnewline
~~     $h_{\texttt{CE}}$&0.185&0.073&0.006&0.012&0.945&0.455&&0.169&0.055&0.009&0.012&0.926&0.491\tabularnewline
\hline
{\bfseries $x=$ 0.5}&&&&&&&&&&&&&\tabularnewline
~~\texttt{locpol}&0.175&0.178&0.007&0.039&1.000&3.929&&0.130&0.106&0.014&0.025&1.000&3.890\tabularnewline
~~\texttt{lpridge}&0.175&0.178&0.007&0.039&0.419&0.326&&\texttt{na}&\texttt{na}&\texttt{na}&\texttt{na}&\texttt{na}&\texttt{na}\tabularnewline
~~\texttt{locfit}&0.175&0.132&0.009&0.026&0.686&0.354&&\texttt{na}&0.428&0.005&0.188&0.000&0.255\tabularnewline
~~\texttt{np}&0.175&0.632&0.004&0.403&0.000&0.256&&0.057&0.100&0.015&0.025&0.762&0.406\tabularnewline
~~\texttt{nprobust}&&&&&&&&&&&&&\tabularnewline
~~      $h_{\texttt{MSE}}$&0.175&0.178&0.007&0.039&0.941&0.468&&0.159&0.150&0.009&0.031&0.943&0.491\tabularnewline
~~     $h_{\texttt{IMSE}}$&0.234&0.295&0.005&0.092&0.931&0.405&&0.182&0.192&0.008&0.045&0.941&0.459\tabularnewline
~~     $h_{\texttt{CE}}$&0.128&0.101&0.010&0.020&0.943&0.547&&0.150&0.134&0.012&0.030&0.938&0.508\tabularnewline
\hline
{\bfseries $x=$ 0.75}&&&&&&&&&&&&&\tabularnewline
~~\texttt{locpol}&0.270&0.095&0.005&0.014&1.000&3.955&&0.130&0.031&0.011&0.012&1.000&3.880\tabularnewline
~~\texttt{lpridge}&0.270&0.095&0.005&0.014&0.705&0.263&&\texttt{na}&\texttt{na}&\texttt{na}&\texttt{na}&\texttt{na}&\texttt{na}\tabularnewline
~~\texttt{locfit}&0.270&0.078&0.005&0.011&0.810&0.287&&\texttt{na}&0.088&0.004&0.012&0.706&0.246\tabularnewline
~~\texttt{np}&0.270&0.045&0.004&0.006&0.838&0.219&&0.057&0.029&0.012&0.013&0.931&0.410\tabularnewline
~~\texttt{nprobust}&&&&&&&&&&&&&\tabularnewline
~~      $h_{\texttt{MSE}}$&0.270&0.095&0.005&0.014&0.938&0.380&&0.159&0.045&0.007&0.009&0.947&0.494\tabularnewline
~~     $h_{\texttt{IMSE}}$&0.234&0.081&0.005&0.012&0.945&0.406&&0.182&0.056&0.007&0.010&0.946&0.461\tabularnewline
~~     $h_{\texttt{CE}}$&0.198&0.064&0.006&0.010&0.947&0.442&&0.169&0.043&0.008&0.010&0.933&0.493\tabularnewline
\hline
{\bfseries $x=$ 1}&&&&&&&&&&&&&\tabularnewline
~~\texttt{locpol}&0.491&0.238&0.020&0.076&1.000&4.108&&0.130&0.001&0.085&0.085&1.000&3.836\tabularnewline
~~\texttt{lpridge}&0.491&0.203&0.018&0.059&0.651&0.509&&\texttt{na}&\texttt{na}&\texttt{na}&\texttt{na}&\texttt{na}&\texttt{na}\tabularnewline
~~\texttt{locfit}&0.491&0.195&0.022&0.060&0.757&0.597&&\texttt{na}&0.225&0.016&0.067&0.556&0.489\tabularnewline
~~\texttt{np}&0.491&0.613&0.011&0.387&0.000&0.225&&0.057&0.006&0.079&0.079&0.701&0.564\tabularnewline
~~\texttt{nprobust}&&&&&&&&&&&&&\tabularnewline
~~      $h_{\texttt{MSE}}$&0.491&0.238&0.020&0.076&0.937&0.783&&0.324&0.010&0.059&0.060&0.895&1.043\tabularnewline
~~     $h_{\texttt{IMSE}}$&0.234&0.042&0.040&0.042&0.936&1.136&&0.182&0.013&0.052&0.052&0.930&1.287\tabularnewline
~~     $h_{\texttt{CE}}$&0.360&0.139&0.026&0.046&0.936&0.915&&0.137&0.006&0.076&0.076&0.917&1.524\tabularnewline
\hline
\end{tabular}
}
	\end{center}
\end{table}

\clearpage
\begin{table}\begin{center}\renewcommand{\arraystretch}{1}
		\caption{Simulation Results, $p=0,deriv=0$\label{table:simuls_p0d0}}\vspace{-.1in}
		\resizebox{\columnwidth}{!}{
\begin{tabular}{lccccccccccccc}
\hline\hline
\multicolumn{1}{l}{\bfseries }&\multicolumn{6}{c}{\bfseries Population bandwidth}&\multicolumn{1}{c}{\bfseries }&\multicolumn{6}{c}{\bfseries Estimated bandwidth}\tabularnewline
\cline{2-7} \cline{9-14}
\multicolumn{1}{l}{}&\multicolumn{1}{c}{$h$}&\multicolumn{1}{c}{Bias}&\multicolumn{1}{c}{Var}&\multicolumn{1}{c}{MSE}&\multicolumn{1}{c}{EC}&\multicolumn{1}{c}{IL}&\multicolumn{1}{c}{}&\multicolumn{1}{c}{$h$}&\multicolumn{1}{c}{Bias}&\multicolumn{1}{c}{Var}&\multicolumn{1}{c}{MSE}&\multicolumn{1}{c}{EC}&\multicolumn{1}{c}{IL}\tabularnewline
\hline
{\bfseries $x=$ 0}&&&&&&&&&&&&&\tabularnewline
~~\texttt{locpol}&\texttt{na}&\texttt{na}&\texttt{na}&\texttt{na}&\texttt{na}&\texttt{na}&&\texttt{na}&\texttt{na}&\texttt{na}&\texttt{na}&\texttt{na}&\texttt{na}\tabularnewline
~~\texttt{lpridge}&0.108&0.086&0.022&0.030&0.911&0.587&&\texttt{na}&\texttt{na}&\texttt{na}&\texttt{na}&\texttt{na}&\texttt{na}\tabularnewline
~~\texttt{locfit}&\texttt{na}&\texttt{na}&\texttt{na}&\texttt{na}&\texttt{na}&\texttt{na}&&\texttt{na}&\texttt{na}&\texttt{na}&\texttt{na}&\texttt{na}&\texttt{na}\tabularnewline
~~\texttt{np}&0.108&0.255&0.010&0.075&0.294&0.396&&0.053&0.093&0.027&0.035&0.870&0.582\tabularnewline
~~\texttt{nprobust}&&&&&&&&&&&&&\tabularnewline
~~      $h_{\texttt{MSE}}$&0.108&0.086&0.022&0.030&0.927&1.122&&0.070&0.048&0.042&0.044&0.919&1.436\tabularnewline
~~     $h_{\texttt{IMSE}}$&0.065&0.049&0.038&0.041&0.913&1.447&&0.119&0.094&0.023&0.032&0.929&1.078\tabularnewline
~~     $h_{\texttt{CE}}$&0.047&0.036&0.052&0.053&0.902&1.674&&0.031&0.029&0.059&0.059&0.897&1.798\tabularnewline
\hline
{\bfseries $x=$ 0.25}&&&&&&&&&&&&&\tabularnewline
~~\texttt{locpol}&\texttt{na}&\texttt{na}&\texttt{na}&\texttt{na}&\texttt{na}&\texttt{na}&&\texttt{na}&\texttt{na}&\texttt{na}&\texttt{na}&\texttt{na}&\texttt{na}\tabularnewline
~~\texttt{lpridge}&0.049&0.007&0.026&0.027&0.942&0.620&&\texttt{na}&\texttt{na}&\texttt{na}&\texttt{na}&\texttt{na}&\texttt{na}\tabularnewline
~~\texttt{locfit}&\texttt{na}&\texttt{na}&\texttt{na}&\texttt{na}&\texttt{na}&\texttt{na}&&\texttt{na}&\texttt{na}&\texttt{na}&\texttt{na}&\texttt{na}&\texttt{na}\tabularnewline
~~\texttt{np}&0.049&0.029&0.012&0.013&0.941&0.435&&0.053&0.033&0.014&0.015&0.936&0.444\tabularnewline
~~\texttt{nprobust}&&&&&&&&&&&&&\tabularnewline
~~      $h_{\texttt{MSE}}$&0.049&0.007&0.026&0.027&0.935&0.619&&0.238&0.108&0.024&0.036&0.765&0.310\tabularnewline
~~     $h_{\texttt{IMSE}}$&0.065&0.011&0.020&0.020&0.936&0.534&&0.119&0.034&0.011&0.012&0.935&0.395\tabularnewline
~~     $h_{\texttt{CE}}$&0.021&0.004&0.057&0.057&0.919&0.913&&0.104&0.019&0.019&0.019&0.890&0.461\tabularnewline
\hline
{\bfseries $x=$ 0.5}&&&&&&&&&&&&&\tabularnewline
~~\texttt{locpol}&\texttt{na}&\texttt{na}&\texttt{na}&\texttt{na}&\texttt{na}&\texttt{na}&&\texttt{na}&\texttt{na}&\texttt{na}&\texttt{na}&\texttt{na}&\texttt{na}\tabularnewline
~~\texttt{lpridge}&0.107&0.072&0.012&0.017&0.892&0.416&&\texttt{na}&\texttt{na}&\texttt{na}&\texttt{na}&\texttt{na}&\texttt{na}\tabularnewline
~~\texttt{locfit}&\texttt{na}&\texttt{na}&\texttt{na}&\texttt{na}&\texttt{na}&\texttt{na}&&\texttt{na}&\texttt{na}&\texttt{na}&\texttt{na}&\texttt{na}&\texttt{na}\tabularnewline
~~\texttt{np}&0.107&0.308&0.006&0.100&0.016&0.294&&0.053&0.087&0.015&0.023&0.814&0.419\tabularnewline
~~\texttt{nprobust}&&&&&&&&&&&&&\tabularnewline
~~      $h_{\texttt{MSE}}$&0.107&0.072&0.012&0.017&0.886&0.415&&0.112&0.076&0.013&0.019&0.846&0.407\tabularnewline
~~     $h_{\texttt{IMSE}}$&0.065&0.028&0.019&0.020&0.930&0.534&&0.119&0.089&0.011&0.019&0.828&0.395\tabularnewline
~~     $h_{\texttt{CE}}$&0.047&0.015&0.026&0.026&0.935&0.631&&0.049&0.016&0.026&0.026&0.931&0.618\tabularnewline
\hline
{\bfseries $x=$ 0.75}&&&&&&&&&&&&&\tabularnewline
~~\texttt{locpol}&\texttt{na}&\texttt{na}&\texttt{na}&\texttt{na}&\texttt{na}&\texttt{na}&&\texttt{na}&\texttt{na}&\texttt{na}&\texttt{na}&\texttt{na}&\texttt{na}\tabularnewline
~~\texttt{lpridge}&0.385&0.156&0.004&0.029&0.266&0.233&&\texttt{na}&\texttt{na}&\texttt{na}&\texttt{na}&\texttt{na}&\texttt{na}\tabularnewline
~~\texttt{locfit}&\texttt{na}&\texttt{na}&\texttt{na}&\texttt{na}&\texttt{na}&\texttt{na}&&\texttt{na}&\texttt{na}&\texttt{na}&\texttt{na}&\texttt{na}&\texttt{na}\tabularnewline
~~\texttt{np}&0.385&0.129&0.003&0.020&0.354&0.217&&0.053&0.027&0.013&0.014&0.936&0.423\tabularnewline
~~\texttt{nprobust}&&&&&&&&&&&&&\tabularnewline
~~      $h_{\texttt{MSE}}$&0.385&0.156&0.004&0.029&0.698&0.240&&0.268&0.060&0.010&0.014&0.818&0.292\tabularnewline
~~     $h_{\texttt{IMSE}}$&0.065&0.009&0.019&0.019&0.939&0.536&&0.119&0.027&0.010&0.011&0.938&0.397\tabularnewline
~~     $h_{\texttt{CE}}$&0.168&0.050&0.007&0.010&0.907&0.332&&0.117&0.018&0.015&0.015&0.905&0.431\tabularnewline
\hline
{\bfseries $x=$ 1}&&&&&&&&&&&&&\tabularnewline
~~\texttt{locpol}&\texttt{na}&\texttt{na}&\texttt{na}&\texttt{na}&\texttt{na}&\texttt{na}&&\texttt{na}&\texttt{na}&\texttt{na}&\texttt{na}&\texttt{na}&\texttt{na}\tabularnewline
~~\texttt{lpridge}&0.186&0.003&0.014&0.014&0.945&0.447&&\texttt{na}&\texttt{na}&\texttt{na}&\texttt{na}&\texttt{na}&\texttt{na}\tabularnewline
~~\texttt{locfit}&\texttt{na}&\texttt{na}&\texttt{na}&\texttt{na}&\texttt{na}&\texttt{na}&&\texttt{na}&\texttt{na}&\texttt{na}&\texttt{na}&\texttt{na}&\texttt{na}\tabularnewline
~~\texttt{np}&0.186&0.168&0.006&0.034&0.428&0.306&&0.053&0.004&0.026&0.026&0.934&0.583\tabularnewline
~~\texttt{nprobust}&&&&&&&&&&&&&\tabularnewline
~~      $h_{\texttt{MSE}}$&0.186&0.003&0.014&0.014&0.936&0.860&&0.123&0.010&0.027&0.027&0.931&1.123\tabularnewline
~~     $h_{\texttt{IMSE}}$&0.065&0.003&0.038&0.038&0.917&1.455&&0.119&0.010&0.022&0.022&0.935&1.080\tabularnewline
~~     $h_{\texttt{CE}}$&0.081&0.006&0.030&0.030&0.923&1.302&&0.054&0.001&0.049&0.049&0.906&1.625\tabularnewline
\hline
\end{tabular}
}
	\end{center}
\end{table}

\clearpage
\begin{table}\begin{center}\renewcommand{\arraystretch}{1}
		\caption{Simulation Results, $p=2,deriv=0$\label{table:simuls_p2d0}}\vspace{-.1in}
		\resizebox{\columnwidth}{!}{
\begin{tabular}{lccccccccccccc}
\hline\hline
\multicolumn{1}{l}{\bfseries }&\multicolumn{6}{c}{\bfseries Population bandwidth}&\multicolumn{1}{c}{\bfseries }&\multicolumn{6}{c}{\bfseries Estimated bandwidth}\tabularnewline
\cline{2-7} \cline{9-14}
\multicolumn{1}{l}{}&\multicolumn{1}{c}{$h$}&\multicolumn{1}{c}{Bias}&\multicolumn{1}{c}{Var}&\multicolumn{1}{c}{MSE}&\multicolumn{1}{c}{EC}&\multicolumn{1}{c}{IL}&\multicolumn{1}{c}{}&\multicolumn{1}{c}{$h$}&\multicolumn{1}{c}{Bias}&\multicolumn{1}{c}{Var}&\multicolumn{1}{c}{MSE}&\multicolumn{1}{c}{EC}&\multicolumn{1}{c}{IL}\tabularnewline
\hline
{\bfseries $x=$ 0}&&&&&&&&&&&&&\tabularnewline
~~\texttt{locpol}&0.442&0.001&0.045&0.045&0.888&0.676&&0.291&0.011&0.078&0.079&0.782&0.674\tabularnewline
~~\texttt{lpridge}&0.442&0.112&0.036&0.049&0.913&0.744&&\texttt{na}&\texttt{na}&\texttt{na}&\texttt{na}&\texttt{na}&\texttt{na}\tabularnewline
~~\texttt{locfit}&0.442&0.014&0.051&0.052&0.950&0.887&&\texttt{na}&0.201&0.034&0.074&0.795&0.705\tabularnewline
~~\texttt{np}&\texttt{na}&\texttt{na}&\texttt{na}&\texttt{na}&\texttt{na}&\texttt{na}&&\texttt{na}&\texttt{na}&\texttt{na}&\texttt{na}&\texttt{na}&\texttt{na}\tabularnewline
~~\texttt{nprobust}&&&&&&&&&&&&&\tabularnewline
~~      $h_{\texttt{MSE}}$&0.442&0.001&0.045&0.045&0.928&1.095&&0.354&0.020&0.086&0.086&0.914&1.264\tabularnewline
~~     $h_{\texttt{IMSE}}$&0.273&0.017&0.075&0.075&0.922&1.399&&0.298&0.021&0.074&0.075&0.925&1.343\tabularnewline
~~     $h_{\texttt{CE}}$&0.254&0.015&0.080&0.081&0.922&1.452&&0.204&0.004&0.130&0.130&0.919&1.681\tabularnewline
\hline
{\bfseries $x=$ 0.25}&&&&&&&&&&&&&\tabularnewline
~~\texttt{locpol}&0.321&0.062&0.008&0.012&1.000&1.653&&0.291&0.046&0.011&0.014&1.000&1.696\tabularnewline
~~\texttt{lpridge}&0.321&0.062&0.008&0.012&0.899&0.359&&\texttt{na}&\texttt{na}&\texttt{na}&\texttt{na}&\texttt{na}&\texttt{na}\tabularnewline
~~\texttt{locfit}&0.321&0.037&0.009&0.010&0.936&0.376&&\texttt{na}&0.132&0.008&0.025&0.651&0.332\tabularnewline
~~\texttt{np}&\texttt{na}&\texttt{na}&\texttt{na}&\texttt{na}&\texttt{na}&\texttt{na}&&\texttt{na}&\texttt{na}&\texttt{na}&\texttt{na}&\texttt{na}&\texttt{na}\tabularnewline
~~\texttt{nprobust}&&&&&&&&&&&&&\tabularnewline
~~      $h_{\texttt{MSE}}$&0.321&0.062&0.008&0.012&0.937&0.366&&0.345&0.083&0.017&0.024&0.912&0.370\tabularnewline
~~     $h_{\texttt{IMSE}}$&0.273&0.031&0.009&0.010&0.941&0.379&&0.298&0.047&0.010&0.012&0.935&0.375\tabularnewline
~~     $h_{\texttt{CE}}$&0.185&0.008&0.013&0.013&0.947&0.457&&0.198&0.012&0.015&0.015&0.945&0.454\tabularnewline
\hline
{\bfseries $x=$ 0.5}&&&&&&&&&&&&&\tabularnewline
~~\texttt{locpol}&0.319&0.082&0.008&0.015&1.000&1.789&&0.291&0.064&0.012&0.016&1.000&1.788\tabularnewline
~~\texttt{lpridge}&0.319&0.082&0.008&0.015&0.846&0.348&&\texttt{na}&\texttt{na}&\texttt{na}&\texttt{na}&\texttt{na}&\texttt{na}\tabularnewline
~~\texttt{locfit}&0.319&0.054&0.009&0.012&0.908&0.371&&\texttt{na}&0.073&0.008&0.014&0.874&0.357\tabularnewline
~~\texttt{np}&\texttt{na}&\texttt{na}&\texttt{na}&\texttt{na}&\texttt{na}&\texttt{na}&&\texttt{na}&\texttt{na}&\texttt{na}&\texttt{na}&\texttt{na}&\texttt{na}\tabularnewline
~~\texttt{nprobust}&&&&&&&&&&&&&\tabularnewline
~~      $h_{\texttt{MSE}}$&0.319&0.082&0.008&0.015&0.842&0.347&&0.290&0.055&0.011&0.014&0.864&0.365\tabularnewline
~~     $h_{\texttt{IMSE}}$&0.273&0.050&0.009&0.012&0.912&0.375&&0.298&0.064&0.010&0.014&0.852&0.362\tabularnewline
~~     $h_{\texttt{CE}}$&0.184&0.014&0.014&0.014&0.939&0.458&&0.167&0.009&0.016&0.016&0.934&0.482\tabularnewline
\hline
{\bfseries $x=$ 0.75}&&&&&&&&&&&&&\tabularnewline
~~\texttt{locpol}&0.401&0.137&0.007&0.026&1.000&1.480&&0.291&0.045&0.012&0.014&1.000&1.696\tabularnewline
~~\texttt{lpridge}&0.401&0.137&0.007&0.026&0.617&0.327&&\texttt{na}&\texttt{na}&\texttt{na}&\texttt{na}&\texttt{na}&\texttt{na}\tabularnewline
~~\texttt{locfit}&0.401&0.087&0.008&0.015&0.837&0.347&&\texttt{na}&0.127&0.008&0.024&0.662&0.332\tabularnewline
~~\texttt{np}&\texttt{na}&\texttt{na}&\texttt{na}&\texttt{na}&\texttt{na}&\texttt{na}&&\texttt{na}&\texttt{na}&\texttt{na}&\texttt{na}&\texttt{na}&\texttt{na}\tabularnewline
~~\texttt{nprobust}&&&&&&&&&&&&&\tabularnewline
~~      $h_{\texttt{MSE}}$&0.401&0.137&0.007&0.026&0.926&0.362&&0.347&0.080&0.016&0.022&0.913&0.370\tabularnewline
~~     $h_{\texttt{IMSE}}$&0.273&0.030&0.009&0.010&0.939&0.379&&0.298&0.046&0.010&0.012&0.932&0.375\tabularnewline
~~     $h_{\texttt{CE}}$&0.231&0.015&0.011&0.011&0.945&0.409&&0.200&0.010&0.015&0.015&0.940&0.454\tabularnewline
\hline
{\bfseries $x=$ 1}&&&&&&&&&&&&&\tabularnewline
~~\texttt{locpol}&0.676&0.288&0.031&0.114&0.616&0.677&&0.291&0.004&0.080&0.080&0.782&0.674\tabularnewline
~~\texttt{lpridge}&0.676&0.205&0.025&0.067&0.741&0.612&&\texttt{na}&\texttt{na}&\texttt{na}&\texttt{na}&\texttt{na}&\texttt{na}\tabularnewline
~~\texttt{locfit}&0.676&0.159&0.034&0.059&0.861&0.724&&\texttt{na}&0.187&0.034&0.069&0.813&0.705\tabularnewline
~~\texttt{np}&\texttt{na}&\texttt{na}&\texttt{na}&\texttt{na}&\texttt{na}&\texttt{na}&&\texttt{na}&\texttt{na}&\texttt{na}&\texttt{na}&\texttt{na}&\texttt{na}\tabularnewline
~~\texttt{nprobust}&&&&&&&&&&&&&\tabularnewline
~~      $h_{\texttt{MSE}}$&0.676&0.288&0.031&0.114&0.902&0.885&&0.357&0.025&0.082&0.083&0.917&1.263\tabularnewline
~~     $h_{\texttt{IMSE}}$&0.273&0.008&0.075&0.075&0.927&1.402&&0.298&0.014&0.074&0.074&0.926&1.345\tabularnewline
~~     $h_{\texttt{CE}}$&0.389&0.014&0.052&0.052&0.932&1.170&&0.206&0.002&0.127&0.127&0.917&1.676\tabularnewline
\hline
\end{tabular}
}
	\end{center}
\end{table}

\clearpage
\begin{table}\begin{center}\renewcommand{\arraystretch}{1}
		\caption{Simulation Results, $p=1,deriv=1$\label{table:simuls_p1d1}}\vspace{-.1in}
		\resizebox{\columnwidth}{!}{
\begin{tabular}{lccccccccccccc}
\hline\hline
\multicolumn{1}{l}{\bfseries }&\multicolumn{6}{c}{\bfseries Population bandwidth}&\multicolumn{1}{c}{\bfseries }&\multicolumn{6}{c}{\bfseries Estimated bandwidth}\tabularnewline
\cline{2-7} \cline{9-14}
\multicolumn{1}{l}{}&\multicolumn{1}{c}{$h$}&\multicolumn{1}{c}{Bias}&\multicolumn{1}{c}{Var}&\multicolumn{1}{c}{MSE}&\multicolumn{1}{c}{EC}&\multicolumn{1}{c}{IL}&\multicolumn{1}{c}{}&\multicolumn{1}{c}{$h$}&\multicolumn{1}{c}{Bias}&\multicolumn{1}{c}{Var}&\multicolumn{1}{c}{MSE}&\multicolumn{1}{c}{EC}&\multicolumn{1}{c}{IL}\tabularnewline
\hline
{\bfseries $x=$ 0}&&&&&&&&&&&&&\tabularnewline
~~\texttt{locpol}&\texttt{na}&\texttt{na}&\texttt{na}&\texttt{na}&\texttt{na}&\texttt{na}&&\texttt{na}&\texttt{na}&\texttt{na}&\texttt{na}&\texttt{na}&\texttt{na}\tabularnewline
~~\texttt{lpridge}&0.228&1.729&2.921&5.911&0.823&6.651&&\texttt{na}&\texttt{na}&\texttt{na}&\texttt{na}&\texttt{na}&\texttt{na}\tabularnewline
~~\texttt{locfit}&\texttt{na}&\texttt{na}&\texttt{na}&\texttt{na}&\texttt{na}&\texttt{na}&&\texttt{na}&\texttt{na}&\texttt{na}&\texttt{na}&\texttt{na}&\texttt{na}\tabularnewline
~~\texttt{np}&0.228&4.357&0.263&19.247&0.000&1.038&&0.057&1.188&32.754&34.166&0.060&0.564\tabularnewline
~~\texttt{nprobust}&&&&&&&&&&&&&\tabularnewline
~~      $h_{\texttt{MSE}}$&0.228&1.729&2.921&5.911&0.937&25.367&&0.369&2.666&5.566&12.674&0.812&20.138\tabularnewline
~~     $h_{\texttt{IMSE}}$&0.146&0.985&11.356&12.326&0.937&49.400&&0.260&2.046&2.992&7.180&0.943&21.715\tabularnewline
~~     $h_{\texttt{CE}}$&0.117&0.745&22.194&22.749&0.932&69.588&&0.190&1.878&24.419&27.946&0.936&54.895\tabularnewline
\hline
{\bfseries $x=$ 0.25}&&&&&&&&&&&&&\tabularnewline
~~\texttt{locpol}&\texttt{na}&\texttt{na}&\texttt{na}&\texttt{na}&\texttt{na}&\texttt{na}&&\texttt{na}&\texttt{na}&\texttt{na}&\texttt{na}&\texttt{na}&\texttt{na}\tabularnewline
~~\texttt{lpridge}&0.405&1.670&0.114&2.903&0.002&1.286&&\texttt{na}&\texttt{na}&\texttt{na}&\texttt{na}&\texttt{na}&\texttt{na}\tabularnewline
~~\texttt{locfit}&\texttt{na}&\texttt{na}&\texttt{na}&\texttt{na}&\texttt{na}&\texttt{na}&&\texttt{na}&\texttt{na}&\texttt{na}&\texttt{na}&\texttt{na}&\texttt{na}\tabularnewline
~~\texttt{np}&0.405&5.280&0.037&27.921&0.000&0.405&&0.057&0.388&4.391&4.541&0.123&0.433\tabularnewline
~~\texttt{nprobust}&&&&&&&&&&&&&\tabularnewline
~~      $h_{\texttt{MSE}}$&0.405&1.670&0.114&2.903&0.031&1.570&&0.213&0.615&0.730&1.108&0.788&2.799\tabularnewline
~~     $h_{\texttt{IMSE}}$&0.146&0.310&1.442&1.539&0.932&4.602&&0.260&0.892&0.435&1.231&0.501&2.091\tabularnewline
~~     $h_{\texttt{CE}}$&0.208&0.631&0.494&0.893&0.844&2.707&&0.109&0.128&4.074&4.090&0.934&7.594\tabularnewline
\hline
{\bfseries $x=$ 0.5}&&&&&&&&&&&&&\tabularnewline
~~\texttt{locpol}&\texttt{na}&\texttt{na}&\texttt{na}&\texttt{na}&\texttt{na}&\texttt{na}&&\texttt{na}&\texttt{na}&\texttt{na}&\texttt{na}&\texttt{na}&\texttt{na}\tabularnewline
~~\texttt{lpridge}&0.147&0.018&1.432&1.433&0.946&4.593&&\texttt{na}&\texttt{na}&\texttt{na}&\texttt{na}&\texttt{na}&\texttt{na}\tabularnewline
~~\texttt{locfit}&\texttt{na}&\texttt{na}&\texttt{na}&\texttt{na}&\texttt{na}&\texttt{na}&&\texttt{na}&\texttt{na}&\texttt{na}&\texttt{na}&\texttt{na}&\texttt{na}\tabularnewline
~~\texttt{np}&0.147&0.064&0.150&0.155&0.891&1.279&&0.057&0.060&4.223&4.226&0.124&0.406\tabularnewline
~~\texttt{nprobust}&&&&&&&&&&&&&\tabularnewline
~~      $h_{\texttt{MSE}}$&0.147&0.018&1.432&1.433&0.947&4.597&&0.436&0.078&0.181&0.187&0.870&1.186\tabularnewline
~~     $h_{\texttt{IMSE}}$&0.146&0.018&1.439&1.439&0.947&4.608&&0.260&0.045&0.334&0.336&0.945&2.038\tabularnewline
~~     $h_{\texttt{CE}}$&0.075&0.121&10.342&10.357&0.942&12.555&&0.224&0.010&0.910&0.910&0.951&3.118\tabularnewline
\hline
{\bfseries $x=$ 0.75}&&&&&&&&&&&&&\tabularnewline
~~\texttt{locpol}&\texttt{na}&\texttt{na}&\texttt{na}&\texttt{na}&\texttt{na}&\texttt{na}&&\texttt{na}&\texttt{na}&\texttt{na}&\texttt{na}&\texttt{na}&\texttt{na}\tabularnewline
~~\texttt{lpridge}&0.182&0.419&0.728&0.904&0.921&3.324&&\texttt{na}&\texttt{na}&\texttt{na}&\texttt{na}&\texttt{na}&\texttt{na}\tabularnewline
~~\texttt{locfit}&\texttt{na}&\texttt{na}&\texttt{na}&\texttt{na}&\texttt{na}&\texttt{na}&&\texttt{na}&\texttt{na}&\texttt{na}&\texttt{na}&\texttt{na}&\texttt{na}\tabularnewline
~~\texttt{np}&0.182&1.699&0.115&3.002&0.000&0.939&&0.057&0.320&4.425&4.527&0.108&0.410\tabularnewline
~~\texttt{nprobust}&&&&&&&&&&&&&\tabularnewline
~~      $h_{\texttt{MSE}}$&0.182&0.419&0.728&0.904&0.920&3.329&&0.219&0.589&0.780&1.128&0.798&2.735\tabularnewline
~~     $h_{\texttt{IMSE}}$&0.146&0.266&1.403&1.473&0.939&4.607&&0.260&0.814&0.458&1.119&0.549&2.094\tabularnewline
~~     $h_{\texttt{CE}}$&0.093&0.132&5.444&5.462&0.947&9.108&&0.113&0.168&3.862&3.890&0.931&7.424\tabularnewline
\hline
{\bfseries $x=$ 1}&&&&&&&&&&&&&\tabularnewline
~~\texttt{locpol}&\texttt{na}&\texttt{na}&\texttt{na}&\texttt{na}&\texttt{na}&\texttt{na}&&\texttt{na}&\texttt{na}&\texttt{na}&\texttt{na}&\texttt{na}&\texttt{na}\tabularnewline
~~\texttt{lpridge}&0.317&1.917&1.072&4.746&0.541&4.040&&\texttt{na}&\texttt{na}&\texttt{na}&\texttt{na}&\texttt{na}&\texttt{na}\tabularnewline
~~\texttt{locfit}&\texttt{na}&\texttt{na}&\texttt{na}&\texttt{na}&\texttt{na}&\texttt{na}&&\texttt{na}&\texttt{na}&\texttt{na}&\texttt{na}&\texttt{na}&\texttt{na}\tabularnewline
~~\texttt{np}&0.317&2.554&0.102&6.626&0.000&0.547&&0.057&0.475&36.754&36.980&0.060&0.564\tabularnewline
~~\texttt{nprobust}&&&&&&&&&&&&&\tabularnewline
~~      $h_{\texttt{MSE}}$&0.317&1.917&1.072&4.746&0.942&15.462&&0.366&1.776&4.913&8.067&0.834&19.592\tabularnewline
~~     $h_{\texttt{IMSE}}$&0.146&0.403&10.763&10.926&0.934&49.521&&0.260&1.324&2.974&4.728&0.944&21.853\tabularnewline
~~     $h_{\texttt{CE}}$&0.163&0.530&7.865&8.146&0.937&42.158&&0.188&1.091&27.014&28.204&0.928&53.489\tabularnewline
\hline
\end{tabular}
}
	\end{center}
\end{table}

\clearpage
\begin{table}\begin{center}\renewcommand{\arraystretch}{1}
		\caption{Simulation Results, $p=2,deriv=1$\label{table:simuls_p2d1}}\vspace{-.1in}
		\resizebox{\columnwidth}{!}{
\begin{tabular}{lccccccccccccc}
\hline\hline
\multicolumn{1}{l}{\bfseries }&\multicolumn{6}{c}{\bfseries Population bandwidth}&\multicolumn{1}{c}{\bfseries }&\multicolumn{6}{c}{\bfseries Estimated bandwidth}\tabularnewline
\cline{2-7} \cline{9-14}
\multicolumn{1}{l}{}&\multicolumn{1}{c}{$h$}&\multicolumn{1}{c}{Bias}&\multicolumn{1}{c}{Var}&\multicolumn{1}{c}{MSE}&\multicolumn{1}{c}{EC}&\multicolumn{1}{c}{IL}&\multicolumn{1}{c}{}&\multicolumn{1}{c}{$h$}&\multicolumn{1}{c}{Bias}&\multicolumn{1}{c}{Var}&\multicolumn{1}{c}{MSE}&\multicolumn{1}{c}{EC}&\multicolumn{1}{c}{IL}\tabularnewline
\hline
{\bfseries $x=$ 0}&&&&&&&&&&&&&\tabularnewline
~~\texttt{locpol}&\texttt{na}&\texttt{na}&\texttt{na}&\texttt{na}&\texttt{na}&\texttt{na}&&\texttt{na}&\texttt{na}&\texttt{na}&\texttt{na}&\texttt{na}&\texttt{na}\tabularnewline
~~\texttt{lpridge}&0.499&3.071&0.808&10.237&0.074&3.490&&\texttt{na}&\texttt{na}&\texttt{na}&\texttt{na}&\texttt{na}&\texttt{na}\tabularnewline
~~\texttt{locfit}&\texttt{na}&\texttt{na}&\texttt{na}&\texttt{na}&\texttt{na}&\texttt{na}&&\texttt{na}&\texttt{na}&\texttt{na}&\texttt{na}&\texttt{na}&\texttt{na}\tabularnewline
~~\texttt{np}&\texttt{na}&\texttt{na}&\texttt{na}&\texttt{na}&\texttt{na}&\texttt{na}&&\texttt{na}&\texttt{na}&\texttt{na}&\texttt{na}&\texttt{na}&\texttt{na}\tabularnewline
~~\texttt{nprobust}&&&&&&&&&&&&&\tabularnewline
~~      $h_{\texttt{MSE}}$&0.499&0.408&3.944&4.110&0.912&19.141&&0.418&0.194&22.266&22.304&0.943&28.538\tabularnewline
~~     $h_{\texttt{IMSE}}$&0.362&0.810&10.852&11.507&0.936&31.135&&0.339&0.781&17.590&18.199&0.951&34.814\tabularnewline
~~     $h_{\texttt{CE}}$&0.350&0.824&12.007&12.686&0.940&32.739&&0.443&0.362&15.837&15.968&0.900&26.414\tabularnewline
\hline
{\bfseries $x=$ 0.25}&&&&&&&&&&&&&\tabularnewline
~~\texttt{locpol}&\texttt{na}&\texttt{na}&\texttt{na}&\texttt{na}&\texttt{na}&\texttt{na}&&\texttt{na}&\texttt{na}&\texttt{na}&\texttt{na}&\texttt{na}&\texttt{na}\tabularnewline
~~\texttt{lpridge}&0.399&1.565&0.168&2.618&0.034&1.576&&\texttt{na}&\texttt{na}&\texttt{na}&\texttt{na}&\texttt{na}&\texttt{na}\tabularnewline
~~\texttt{locfit}&\texttt{na}&\texttt{na}&\texttt{na}&\texttt{na}&\texttt{na}&\texttt{na}&&\texttt{na}&\texttt{na}&\texttt{na}&\texttt{na}&\texttt{na}&\texttt{na}\tabularnewline
~~\texttt{np}&\texttt{na}&\texttt{na}&\texttt{na}&\texttt{na}&\texttt{na}&\texttt{na}&&\texttt{na}&\texttt{na}&\texttt{na}&\texttt{na}&\texttt{na}&\texttt{na}\tabularnewline
~~\texttt{nprobust}&&&&&&&&&&&&&\tabularnewline
~~      $h_{\texttt{MSE}}$&0.399&1.565&0.169&2.617&0.922&2.812&&0.331&1.143&0.343&1.649&0.916&3.998\tabularnewline
~~     $h_{\texttt{IMSE}}$&0.362&1.472&0.170&2.337&0.930&3.214&&0.339&1.356&0.207&2.045&0.939&3.512\tabularnewline
~~     $h_{\texttt{CE}}$&0.280&1.083&0.216&1.388&0.944&4.277&&0.272&0.976&0.420&1.372&0.934&5.145\tabularnewline
\hline
{\bfseries $x=$ 0.5}&&&&&&&&&&&&&\tabularnewline
~~\texttt{locpol}&\texttt{na}&\texttt{na}&\texttt{na}&\texttt{na}&\texttt{na}&\texttt{na}&&\texttt{na}&\texttt{na}&\texttt{na}&\texttt{na}&\texttt{na}&\texttt{na}\tabularnewline
~~\texttt{lpridge}&0.995&0.187&0.027&0.062&0.765&0.611&&\texttt{na}&\texttt{na}&\texttt{na}&\texttt{na}&\texttt{na}&\texttt{na}\tabularnewline
~~\texttt{locfit}&\texttt{na}&\texttt{na}&\texttt{na}&\texttt{na}&\texttt{na}&\texttt{na}&&\texttt{na}&\texttt{na}&\texttt{na}&\texttt{na}&\texttt{na}&\texttt{na}\tabularnewline
~~\texttt{np}&\texttt{na}&\texttt{na}&\texttt{na}&\texttt{na}&\texttt{na}&\texttt{na}&&\texttt{na}&\texttt{na}&\texttt{na}&\texttt{na}&\texttt{na}&\texttt{na}\tabularnewline
~~\texttt{nprobust}&&&&&&&&&&&&&\tabularnewline
~~      $h_{\texttt{MSE}}$&0.995&0.187&0.027&0.062&0.942&1.524&&0.525&0.086&0.078&0.086&0.958&2.042\tabularnewline
~~     $h_{\texttt{IMSE}}$&0.362&0.075&0.091&0.097&0.945&2.786&&0.339&0.067&0.119&0.124&0.948&3.137\tabularnewline
~~     $h_{\texttt{CE}}$&0.697&0.177&0.027&0.059&0.942&1.538&&0.260&0.023&0.356&0.356&0.946&5.169\tabularnewline
\hline
{\bfseries $x=$ 0.75}&&&&&&&&&&&&&\tabularnewline
~~\texttt{locpol}&\texttt{na}&\texttt{na}&\texttt{na}&\texttt{na}&\texttt{na}&\texttt{na}&&\texttt{na}&\texttt{na}&\texttt{na}&\texttt{na}&\texttt{na}&\texttt{na}\tabularnewline
~~\texttt{lpridge}&0.408&1.470&0.165&2.327&0.050&1.574&&\texttt{na}&\texttt{na}&\texttt{na}&\texttt{na}&\texttt{na}&\texttt{na}\tabularnewline
~~\texttt{locfit}&\texttt{na}&\texttt{na}&\texttt{na}&\texttt{na}&\texttt{na}&\texttt{na}&&\texttt{na}&\texttt{na}&\texttt{na}&\texttt{na}&\texttt{na}&\texttt{na}\tabularnewline
~~\texttt{np}&\texttt{na}&\texttt{na}&\texttt{na}&\texttt{na}&\texttt{na}&\texttt{na}&&\texttt{na}&\texttt{na}&\texttt{na}&\texttt{na}&\texttt{na}&\texttt{na}\tabularnewline
~~\texttt{nprobust}&&&&&&&&&&&&&\tabularnewline
~~      $h_{\texttt{MSE}}$&0.408&1.468&0.166&2.321&0.921&2.729&&0.387&1.140&0.325&1.626&0.877&3.540\tabularnewline
~~     $h_{\texttt{IMSE}}$&0.362&1.368&0.168&2.041&0.932&3.214&&0.339&1.260&0.203&1.791&0.939&3.513\tabularnewline
~~     $h_{\texttt{CE}}$&0.286&1.036&0.206&1.280&0.942&4.187&&0.295&0.981&0.388&1.350&0.925&4.721\tabularnewline
\hline
{\bfseries $x=$ 1}&&&&&&&&&&&&&\tabularnewline
~~\texttt{locpol}&\texttt{na}&\texttt{na}&\texttt{na}&\texttt{na}&\texttt{na}&\texttt{na}&&\texttt{na}&\texttt{na}&\texttt{na}&\texttt{na}&\texttt{na}&\texttt{na}\tabularnewline
~~\texttt{lpridge}&0.486&2.171&0.863&5.576&0.342&3.613&&\texttt{na}&\texttt{na}&\texttt{na}&\texttt{na}&\texttt{na}&\texttt{na}\tabularnewline
~~\texttt{locfit}&\texttt{na}&\texttt{na}&\texttt{na}&\texttt{na}&\texttt{na}&\texttt{na}&&\texttt{na}&\texttt{na}&\texttt{na}&\texttt{na}&\texttt{na}&\texttt{na}\tabularnewline
~~\texttt{np}&\texttt{na}&\texttt{na}&\texttt{na}&\texttt{na}&\texttt{na}&\texttt{na}&&\texttt{na}&\texttt{na}&\texttt{na}&\texttt{na}&\texttt{na}&\texttt{na}\tabularnewline
~~\texttt{nprobust}&&&&&&&&&&&&&\tabularnewline
~~      $h_{\texttt{MSE}}$&0.486&0.012&4.336&4.336&0.925&19.944&&0.434&0.132&21.676&21.694&0.931&27.815\tabularnewline
~~     $h_{\texttt{IMSE}}$&0.362&0.819&10.611&11.281&0.940&31.154&&0.339&0.735&17.488&18.028&0.949&34.841\tabularnewline
~~     $h_{\texttt{CE}}$&0.341&0.817&12.662&13.330&0.938&34.079&&0.446&0.290&14.967&15.052&0.903&26.146\tabularnewline
\hline
\end{tabular}
}
	\end{center}
\end{table}

\begin{table}
	\renewcommand{\arraystretch}{1.3}
	\begin{center}
		\caption{Bandwidth Selection Methods with Simulated Data\label{table:simuls_bw}}\vspace{-.1in}
		\subfloat[$p=1,deriv=0$]{\resizebox{!}{.9in}{
\begin{tabular}{lccccccc}
\hline\hline
\multicolumn{1}{l}{\bfseries }&\multicolumn{3}{c}{\bfseries MSE}&\multicolumn{1}{c}{\bfseries }&\multicolumn{3}{c}{\bfseries CE}\tabularnewline
\cline{2-4} \cline{6-8}
\multicolumn{1}{l}{}&\multicolumn{1}{c}{POP}&\multicolumn{1}{c}{DPI}&\multicolumn{1}{c}{ROT}&\multicolumn{1}{c}{}&\multicolumn{1}{c}{POP}&\multicolumn{1}{c}{DPI}&\multicolumn{1}{c}{ROT}\tabularnewline
\hline
&&&&&&&\tabularnewline
$x= 0 $&0.347&0.298&0.152&&0.254&0.135&0.218\tabularnewline
$x= 0.25 $&0.253&0.163&0.673&&0.185&0.169&0.120\tabularnewline
$x= 0.5 $&0.175&0.159&0.209&&0.128&0.150&0.117\tabularnewline
$x= 0.75 $&0.270&0.159&0.732&&0.198&0.169&0.117\tabularnewline
$x= 1 $&0.491&0.324&0.155&&0.360&0.137&0.238\tabularnewline
\hline
&&&&&&&\tabularnewline
\textbf{IMSE}&0.234&0.182&0.208&&&&\tabularnewline
\hline
\end{tabular}
}}\bigskip\\
		\subfloat[$p=0,deriv=0$]{\resizebox{.45\columnwidth}{1in}{
\begin{tabular}{lccccccc}
\hline\hline
\multicolumn{1}{l}{\bfseries }&\multicolumn{3}{c}{\bfseries MSE}&\multicolumn{1}{c}{\bfseries }&\multicolumn{3}{c}{\bfseries CE}\tabularnewline
\cline{2-4} \cline{6-8}
\multicolumn{1}{l}{}&\multicolumn{1}{c}{POP}&\multicolumn{1}{c}{DPI}&\multicolumn{1}{c}{ROT}&\multicolumn{1}{c}{}&\multicolumn{1}{c}{POP}&\multicolumn{1}{c}{DPI}&\multicolumn{1}{c}{ROT}\tabularnewline
\hline
&&&&&&&\tabularnewline
$x= 0 $&0.108&0.070&0.093&&0.047&0.031&0.031\tabularnewline
$x= 0.25 $&0.049&0.238&0.061&&0.021&0.104&0.104\tabularnewline
$x= 0.5 $&0.107&0.112&0.128&&0.047&0.049&0.049\tabularnewline
$x= 0.75 $&0.385&0.268&0.107&&0.168&0.117&0.117\tabularnewline
$x= 1 $&0.186&0.123&0.155&&0.081&0.054&0.054\tabularnewline
\hline
&&&&&&&\tabularnewline
\textbf{IMSE}&0.065&0.119&0.078&&&&\tabularnewline
\hline
\end{tabular}
}}\qquad
		\subfloat[$p=1,deriv=1$]{\resizebox{.45\columnwidth}{1in}{
\begin{tabular}{lccccccc}
\hline\hline
\multicolumn{1}{l}{\bfseries }&\multicolumn{3}{c}{\bfseries MSE}&\multicolumn{1}{c}{\bfseries }&\multicolumn{3}{c}{\bfseries CE}\tabularnewline
\cline{2-4} \cline{6-8}
\multicolumn{1}{l}{}&\multicolumn{1}{c}{POP}&\multicolumn{1}{c}{DPI}&\multicolumn{1}{c}{ROT}&\multicolumn{1}{c}{}&\multicolumn{1}{c}{POP}&\multicolumn{1}{c}{DPI}&\multicolumn{1}{c}{ROT}\tabularnewline
\hline
&&&&&&&\tabularnewline
$x= 0 $&0.228&0.369&0.124&&0.117&0.190&0.190\tabularnewline
$x= 0.25 $&0.405&0.213&0.476&&0.208&0.109&0.109\tabularnewline
$x= 0.5 $&0.147&0.436&0.170&&0.075&0.224&0.224\tabularnewline
$x= 0.75 $&0.182&0.219&0.510&&0.093&0.113&0.113\tabularnewline
$x= 1 $&0.317&0.366&0.127&&0.163&0.188&0.188\tabularnewline
\hline
&&&&&&&\tabularnewline
\textbf{IMSE}&0.146&0.260&0.171&&&&\tabularnewline
\hline
\end{tabular}
}}\bigskip\\
		\subfloat[$p=2,deriv=0$]{\resizebox{.45\columnwidth}{1in}{
\begin{tabular}{lccccccc}
\hline\hline
\multicolumn{1}{l}{\bfseries }&\multicolumn{3}{c}{\bfseries MSE}&\multicolumn{1}{c}{\bfseries }&\multicolumn{3}{c}{\bfseries CE}\tabularnewline
\cline{2-4} \cline{6-8}
\multicolumn{1}{l}{}&\multicolumn{1}{c}{POP}&\multicolumn{1}{c}{DPI}&\multicolumn{1}{c}{ROT}&\multicolumn{1}{c}{}&\multicolumn{1}{c}{POP}&\multicolumn{1}{c}{DPI}&\multicolumn{1}{c}{ROT}\tabularnewline
\hline
&&&&&&&\tabularnewline
$x= 0 $&0.442&0.354&0.411&&0.254&0.204&0.204\tabularnewline
$x= 0.25 $&0.321&0.345&0.369&&0.185&0.198&0.198\tabularnewline
$x= 0.5 $&0.319&0.290&0.672&&0.184&0.167&0.167\tabularnewline
$x= 0.75 $&0.401&0.347&0.376&&0.231&0.200&0.200\tabularnewline
$x= 1 $&0.676&0.357&0.416&&0.389&0.206&0.206\tabularnewline
\hline
&&&&&&&\tabularnewline
\textbf{IMSE}&0.273&0.298&0.371&&&&\tabularnewline
\hline
\end{tabular}
}}\qquad
		\subfloat[$p=2,deriv=1$]{\resizebox{.45\columnwidth}{1in}{
\begin{tabular}{lccccccc}
\hline\hline
\multicolumn{1}{l}{\bfseries }&\multicolumn{3}{c}{\bfseries MSE}&\multicolumn{1}{c}{\bfseries }&\multicolumn{3}{c}{\bfseries CE}\tabularnewline
\cline{2-4} \cline{6-8}
\multicolumn{1}{l}{}&\multicolumn{1}{c}{POP}&\multicolumn{1}{c}{DPI}&\multicolumn{1}{c}{ROT}&\multicolumn{1}{c}{}&\multicolumn{1}{c}{POP}&\multicolumn{1}{c}{DPI}&\multicolumn{1}{c}{ROT}\tabularnewline
\hline
&&&&&&&\tabularnewline
$x= 0 $&0.499&0.418&0.401&&0.350&0.443&0.293\tabularnewline
$x= 0.25 $&0.399&0.331&0.352&&0.280&0.272&0.232\tabularnewline
$x= 0.5 $&0.995&0.525&0.696&&0.697&0.260&0.368\tabularnewline
$x= 0.75 $&0.408&0.387&0.359&&0.286&0.295&0.271\tabularnewline
$x= 1 $&0.486&0.434&0.406&&0.341&0.446&0.304\tabularnewline
\hline
&&&&&&&\tabularnewline
\textbf{IMSE}&0.362&0.339&0.352&&&&\tabularnewline
\hline
\end{tabular}
}}
	\end{center}
\end{table}

\clearpage

\end{document}